\documentclass[runningheads,10pt]{llncs}
\usepackage[T1]{fontenc}

\usepackage{amsmath}
\usepackage{amsfonts}
\usepackage{mathrsfs}
\usepackage{booktabs}
\usepackage{multirow}
\usepackage{subfigure}
\usepackage[hidelinks]{hyperref}
\usepackage{url}
\usepackage{graphicx}
\usepackage{textcomp}
\usepackage{xcolor}
\usepackage{float}
\usepackage{caption}
\usepackage{bbding}
\usepackage{cite}
\begin{document}
\sloppy

\title{Membership Inference of Diffusion Models}

\titlerunning{Membership Inference of Diffusion Models}

\author{Hailong Hu\inst{1}  \and
	Jun Pang\inst{2,1}}
\authorrunning{Hu and Pang}
\institute{SnT, University of Luxembourg, Esch-sur-Alzette, Luxembourg  \and
	FSTM, University of Luxembourg, Esch-sur-Alzette, Luxembourg 
}

\maketitle             

\begin{abstract}
Recent years have witnessed the tremendous success of diffusion models in data synthesis. 
However, when diffusion models are applied to sensitive data, they also give rise to severe privacy concerns.
In this paper, 
we systematically present the first study about membership inference attacks against diffusion models, which aims to infer whether a sample was used to train the model.
Two attack methods are proposed, namely loss-based and likelihood-based attacks.
Our attack methods are evaluated on several state-of-the-art diffusion models, over different datasets in relation to privacy-sensitive data. 
Extensive experimental evaluations show that our attacks can achieve remarkable performance.
Furthermore, we exhaustively investigate various factors which can affect attack performance.
Finally, we also evaluate the performance of our attack methods on diffusion models trained with differential privacy. 
	
\end{abstract}

\thispagestyle{plain}
\pagestyle{plain}

\section{Introduction}
\label{sec:intro}
Diffusion models~\cite{sohl2015deep} have recently made remarkable progress in image synthesis~\cite{ho2020denoising,song2020score,karras2022elucidating}, even being able to generate better-quality images than generative adversarial networks~(GANs)~\cite{goodfellow2014generative} in some situations~\cite{dhariwal2021diffusion}. 
They have also been applied to sensitive personal data, such as the human face~\cite{song2019generative,karras2022elucidating} or medical images~\cite{pinaya2022brain,kazerouni2022diffusion}, which might unwittingly lead to the leakage of training data.
As a consequence, it is paramount to study privacy breaches in diffusion models.

Membership inference~(MI) attacks aim to infer whether a given sample was used to train the model~\cite{shokri2017membership}.
In practice, they are widely applied to analyze the privacy risks of a machine learning model~\cite{INPTL,murakonda2020ml}.
To date, a growing number of studies concentrate on classification models~\cite{salemml,carlini2021membership,shokri2017membership,ye2021enhanced,liu2022membership}, GANs~\cite{chen2019gan,hayes2019logan}, text-to-image generative models~\cite{wu2022membership}, and language models~\cite{carlini2019secret,carlini2020extracting}.
However, there is still a lack of work on MI attacks against diffusion models.
In addition, data protection regulations, such as GDPR~\cite{GDPR}, require that it is mandatory to assess privacy threats of technologies when they are involving sensitive data.
Therefore, all of these drive us to investigate the membership vulnerability of diffusion models.

In this paper, we systematically study the problem of membership inference of diffusion models. 
Specifically, we consider two threat models: in threat model~I, adversaries are allowed to obtain the target diffusion model, which  means that adversaries can calculate the loss values of a sample through the model.
This scenario usually occurs when institutions share a generative model with their collaborators to avoid directly sharing original data~\cite{park2018data,lin2020using}.
We emphasize that obtaining losses of a model is realistic because it is widely adopted in studying MI attacks on classification models~\cite{carlini2021membership,shokri2017membership,ye2021enhanced,liu2022membership}.
In threat model II, adversaries can obtain the likelihood value of a sample from a diffusion model.
Providing the exact likelihood value of any sample is one of the advantages of diffusion models~\cite{song2020score}.
Thus, here we aim to study whether the likelihood value of a sample can be considered as a clue to infer membership.  
Based on both threat models, two types of attack methods are developed respectively: loss-based attack and likelihood-based attack. They are detailed in  Section~\ref{sec:Methodology}.

We evaluate our methods on four state-of-the-art diffusion models: DDPM~\cite{ho2020denoising}, SMLD~\cite{song2019generative}, VPSDE~\cite{song2020score} and VESDE~\cite{song2020score}.
We use two privacy-sensitive datasets: a human face dataset FFHQ~\cite{StyleGAN12019style} and a diabetic retinopathy dataset DRD~\cite{DRD}.
Extensive experimental evaluations show that our attack methods can
achieve superior attack performance~(see Section~\ref{sec:Evaluation}).
For instance, on the target model DDPM trained on FFHQ-1k, our loss-based attack can achieve a $100\%$ true positive rate~(TPR) even at $0.01\%$ false negative rate~(FPR) when the diffusion step is 200. 
Similar performance can be also seen on our likelihood-based attack where $71\%$ TPR at $0.01\%$ FPR can be obtained, which is 7,100 times more powerful than random guesses.
We also analyze attack performance with respect to various factors~(see Section~\ref{sec:Analysis}).
For example, we find that both attacks gradually become weak with the increase in the number of training samples. 
In addition, similar to FFHQ, both attacks can still achieve excellent performance on the medical dataset DRD.
Finally, we also evaluate our attack performance on a classical defense --- differential privacy~\cite{dwork2008differential}~(see Section~\ref{sec:Defenses}). 
Specifically, we train target models using differentially-private stochastic gradient descent (DP-SGD)~\cite{abadi2016deep}.
Extensive evaluations show that although the performance of both types of attack can be alleviated on models trained with DP-SGD, they sacrifice too much model utility, which also gives a new research direction for the future.

Our contributions can be summarized as follows:
(1) we perform the first study of MI attacks against diffusion models; (2) 
we propose two types of attacks to infer membership of diffusion models, showing that different diffusion steps of a diffusion model have significantly different privacy risks and the likelihood values of samples from a diffusion model are a strong clue to infer training samples;
(3) we evaluate our attacks on one classical defense --- diffusion models trained with DP-SGD, finding that it mitigates our attacks at the cost of the quality of synthetic samples.
In the end, we want to emphasize that although we study membership inference from the perspective of attackers, our proposed methods can directly be applied to audit the privacy risks of diffusion models when model providers need to evaluate the privacy risks of their models.

\section{Background: Diffusion Models}
\label{sec:Background}

Diffusion models~\cite{sohl2015deep} are a class of probabilistic generative models. 
They aim to learn the distribution of a training set, and the resulting model can be utilized to synthesize new data samples~\cite{StatisticalMechanics}.

In general, a diffusion model includes two processes: a forward process and a reverse process~\cite{sohl2015deep}.
In the forward process, i.e. the diffusion process, 
it aims to transform a complex data distribution~$p_{\it data}$ into a simple prior distribution, e.g. Gaussian distribution~$\mathcal{N}(0,\,\sigma^{2}\mathrm{I})$, by gradually adding different levels of noise~$0=\sigma_{0}<\sigma_{1}<,...,<\sigma_{T}=\sigma_{\it max}$, into the data $x$.
In the reverse process, 
it targets at synthesizing a new data sample~$\tilde x_{0}$ through step by step denoising a data sample~$\tilde x_{T} \sim \mathcal{N}(0,\,\sigma_{max}^{2}\mathrm{I})$.
Both processes are defined as Markov chains, and the transitions from one step to another step are described by transition kernels.
In the following, we briefly introduce three typical diffusion models.

\smallskip\noindent
\textbf{DDPM.} A denoising diffusion probabilistic model~(DDPM) proposed by Ho et al.~\cite{ho2020denoising} defines the forward process: $q(x_1,...,x_T|x_0)=\prod_{t=1}^{T}q(x_t|x_{t-1})$,where $T$ is the number of diffusion steps.
The transition kernel uses a Gaussian transition kernel: $q(x_t|x_{t-1})=\mathcal{N}(x_t;\sqrt{1-\beta_t}x_{t-1},\beta_t\mathrm{I})$, where the hyperparameter $\beta_t \in (0,1)$ is a variance schedule. 
Based on the transition kernel, we can get a perturbed sample by: $x_t\gets \sqrt{1-\beta_{t}}x_{t-1}+\sqrt{\beta_{t}}\varepsilon$, where $\varepsilon\sim \mathcal{N}(0,\mathrm{I})$.
The transition kernel from the initial step to any $t$ step can be expressed as:~$	q(x_t|x_0)=\mathcal{N}(x_t;\sqrt[]{\bar{\alpha_t}}x_0,(1-\bar{\alpha_t})\mathrm{I})$, where  $\bar{\alpha_t}=\prod_{i=0}^{t}\alpha_i$ and $\alpha_t:=1-\beta_t$.
Therefore, any perturbed sample can be obtained by: $x_t\gets \sqrt{\bar{\alpha_t}}x_0+\sqrt{1-\bar{\alpha_t}}\varepsilon$.
In the reverse process, DDPM generates a new sample by: $\tilde{x}_{t-1}\gets \frac{1}{\sqrt[]{{\alpha_t}}}(\tilde{x}_t-\frac{\beta_t}{{\sqrt[]{1-{\bar \alpha_{t}}}}}\mathrm{\varepsilon}_\theta(\tilde x_t,t))+\sigma_t\epsilon$,
where $\epsilon_\theta(x_t,t)$ is a neural network predicting noise.
In practice, DDPM is trained by minimizing the following loss:
\begin{equation}
L(\theta) = \mathbb{E}_{t \sim [1,T],x \sim p_{\it data},\varepsilon\sim \mathcal{N}(0,\mathrm{I})}[||\varepsilon - \varepsilon_\theta(\sqrt{\bar{\alpha_t}}x + \sqrt{1-\bar{\alpha_t}}\varepsilon ,t)||^2]. 
\label{eq:ddpm}
\end{equation}

\smallskip\noindent
\textbf{SMLD.} Score matching with Langevin dynamics~(SMLD)~\cite{song2019generative} first learns to estimate the \textit{score}, then generates new samples by Langevin dynamics.
The \textit{score} refers to the gradient of the log probability density with respect to data, i.e. $\nabla_x\text{\it log}\: p(x)$. 
The transition kernel in the forward process is: $q(x_t|x_0) = \mathcal{N}(x_t;x_0,\sigma_t^2\mathrm{I})$.
Thus, a perturbed sample is obtained by: $x_t\gets x_0+\sigma_t\epsilon$.
In the reverse process, SMLD uses an annealed Langevin dynamics to generate a new sample by: $\tilde{x}_t\gets \tilde{x}_{t-1}+\frac{\alpha_i}{2}s_\theta(\tilde{x}_{t-1},\sigma_i)+\sqrt{\alpha_i}\epsilon$, where the hyperparameter $\sigma_i$ controls the updating magnitudes and $s_\theta(x_t,\sigma_i)$ is a noise conditioned neural network predicting the \textit{score}.
Training of the SMLD is performed by minimizing the following loss:
\begin{equation}
L_\theta=\mathbb{E}_{t\sim[1,T],x \sim p_{\it data},x_t \sim q_{\sigma_t}(x_t|x)}[\lambda(\sigma_t)||s_\theta(x_t,\sigma_t) -\nabla_{x_t} \text{log}\:q_{\sigma_t}(x_t|x)||^2],
\label{eq:SMLD}
\end{equation}
where $\lambda(\sigma_t)$ is a coefficient function and $\nabla_{x_t}\text{log}\:q_{\sigma_t}(x_t|x)=-\frac{x_t-x}{\sigma_t^2}$,

\smallskip\noindent
\textbf{SSDE.} Unlike prior works DDPM or SMLD which utilize a finite number of noise distributions, i.e. $t$ is discrete and usually at most $T$, Song et al.~\cite{song2020score} propose a score-based generative framework through the lens of stochastic differential equations~(SDEs), which can add an infinite number of noise distributions to further improve the performance of generative models.
The forward process which adds an infinite number of noise distributions can be described as a continuous-time stochastic process.
Specifically, the forward process of the score-based SDE~(SSDE) is defined as:
\begin{equation}
\text{d}x=\text{f}(x,t)dt+g(t)\text{d}{w},
\label{eq:ssde_forward}
\end{equation}
where $\text{f}(x,t)$, $g(t)$ and $d\text{w}$ are the drift coefficient, the diffusion coefficient and a standard Wiener process, respectively.
The reverse process corresponds to a reverse-time SDE:~$dx=[\text{f}(x,t)-g(t)^2\nabla _x\text{log}~q_t(x)]\text{d}t+g(t)\text{d}\bar{w}$, where $\bar{w}$ is a standard Wiener process in the reverse time.
Training of the SSDE is performed by minimizing the following loss:
\begin{equation}
L_\theta=\mathbb{E}_{t\in \mathcal{U}(0,T),x \sim p_{\it data},x_t \sim q(x_t|x)}[\lambda(t)||s_\theta(x_t,t)-\nabla_{x_t}\text{log}~q(x_t|x)||^2].
\label{eq:SSDE}
\end{equation}
The SSDE is a general and unified framework.
Based on different coefficients in Equation~\ref{eq:ssde_forward}, the variance preserving~(VP) and variance exploding (VE) are instantiated. 
The VPSDE is defined as: $\text{d}x=-\frac{1}{2}\beta(t)x\text{d}t+\sqrt{\beta (t)}\text{d}w$.
The VESDE is defined as: $dx=\sqrt{\frac{\text{d}[\sigma^2(t)]}{\text{d}t}}\text{d}w$.
Furthermore, the SSDE also shows the noise perturbations of DDPM and SMLD are discretizations of VP and VE, respectively.
Note that, diffusion steps usually used in diffusion models also refer to time steps that are used in SDEs. 
In this work, we study the privacy risks of four target models: DDPM, SMLD, VPSDE, and VESDE.

\section{Methodology}
\label{sec:Methodology}
The objective of MI attacks is to infer if a sample was used to train a model.
This provides model providers with a method to evaluate the information leakage of a machine learning model.
In this section, we first introduce threat models and then present our MI methods.

\subsection{Threat Models}
\smallskip\noindent
\textbf{Threat Model I.} In this setting, we assume adversaries can only obtain the target model, i.e. the victim diffusion model.
This setting is realistic because institutions are prone to share generative models with their collaborators instead of directly utilizing original data, considering privacy threats or data regulations~\cite{park2018data,lin2020using}.
We emphasize that adversaries do not gain any knowledge of the training set.
Obtaining the target model indicates that adversaries can get the loss values through the model, and this is realistic because most MI attacks on classification models also assume adversaries can get loss values~\cite{carlini2021membership,shokri2017membership,ye2021enhanced,liu2022membership}.
Under this threat model, we propose a loss-based MI attack.

\smallskip\noindent
\textbf{Threat Model II.} In this setting, we assume adversaries are able to have access to the likelihood values of samples from a diffusion model. 
Diffusion models have advantages in providing the exact likelihood value of any sample~\cite{song2020score}.
Here we aim to study whether the likelihood values of samples can be utilized as a signal to perform membership inference.
Under this threat model, we propose a likelihood-based MI attack.

\subsection{Attack Methods}
\smallskip\noindent
\textbf{Problem Formulation.}
Given a target diffusion model~$G_{\it tar}$, the objective of membership inference attacks is to infer whether a sample~$x$ from a target dataset~$X_{\it tar}$ is used to train the~$G_{\it tar}$.

\smallskip\noindent
\textbf{Loss-based Attack.}
For threat model I, we develop a loss-based attack.
As introduced in Section~\ref{sec:Background}, diffusion models can add an infinite or finite number of noise distributions, which are corresponding to continuous or discrete SDE, respectively.
Therefore, we can calculate the loss value of a sample at each diffusion step $t$.
Specifically, based on Equation~\ref{eq:ddpm}, the loss of a sample $x$ at $t$ diffusion step of DDPM is calculated by: 
\begin{equation}
L = \frac{1}{m}\sum_{}^{}||\varepsilon - \varepsilon_{\theta^\star}(\sqrt{\bar{\alpha_t}}x + \sqrt{1-\bar{\alpha_t}}\varepsilon ,t)||^2, 
\end{equation}
where $m$ is the dimension of $x$ and $\varepsilon_{\theta^\star}(.)$ is the trained network. By Equation~\ref{eq:SMLD}, the loss of a sample $x$ at $t$ diffusion step of SMLD is calculated by: 
\begin{equation}
L = \frac{1}{m}\sum_{}^{}\lambda(\sigma_t)||s_{\theta^\star}(x_t,\sigma_t) -\nabla_{x_t} \text{log}\:q_{\sigma_t}(x_t|x)||^2,
\end{equation}
where $s_{\theta^\star}(.)$ is the trained network.
Based on Equation~\ref{eq:SSDE}, the loss of a sample $x$ at $t$ diffusion step of VPSDE and VESDE is: 
\begin{equation}
L = \frac{1}{m}\sum_{}^{}\lambda(t)||s_{\theta^\star}(x_t,t)-\nabla_{x_t}\text{log}~q(x_t|x)||^2.
\end{equation}
Then, we make a membership inference directly based on the loss value of a sample at one diffusion step.
Namely, if a sample’s loss value is less than certain thresholds, this sample is
marked as a member sample.
For one sample, we can get $T$ or infinite losses, depending on continuous or discrete SDEs.
In this work, in order to thoroughly demonstrate the performance of our attack, 
we compute losses of all diffusion steps $T$ for the discrete case.
We randomly select $T$ diffusion steps for the continuous case although it has infinite steps.

\smallskip\noindent
\textbf{Likelihood-based Attack.}
For threat model II, we propose to utilize the likelihood of a sample to infer membership.
We compute the log-likelihood of a sample $x$ based on the following equation proposed by~\cite{song2020score}.
\begin{equation}
\text{log}~p(x)=\text{log}~p_T(x_T)-\int_{0}^{T}\nabla \cdot \tilde{\mathrm{f}}_{\theta^\star}(x_t,t)dt,
\label{eq:est_likelih}
\end{equation}
where $\nabla \cdot \tilde{\mathrm{f}}_{\theta^\star}(x,t)$ is estimated by the Skilling-Hutchinson trace estimator~\cite{grathwohl2018ffjord}. If the log-likelihood value of a sample is higher than certain thresholds, this sample is
predicted as a member sample.
As introduced in Section~\ref{sec:Background}, the work SSDE~\cite{song2020score} is a unified framework.
In other words, DDPM, SMLD, VPSDE and VESDE can be described by Equation~\ref{eq:ssde_forward}.
Therefore, Equation~\ref{eq:est_likelih} can be applied to these models to estimate the likelihood of one sample.
In this work, we compute the likelihood values of all training samples.

\section{Experiments}
\label{sec:Experiments}

\subsection{Datasets}
\label{ssec:Datasets}
We use two different datasets to evaluate our attack methods. They cover the human face and medical images, which are all considered privacy-sensitive data. 

\smallskip\noindent
\textbf{FFHQ.} 
The Flickr-Faces-HQ dataset~(FFHQ)~\cite{StyleGAN12019style} is a new dataset that contains $70,000$ high-quality human face images. 
In this work, we randomly choose $1,000$ images to train target models.
We also explore the effect of the size of the training set in Section~\ref{ssec:effects_size_data}.

\smallskip\noindent
\textbf{DRD.}
The Diabetic Retinopathy dataset~(DRD)~\cite{DRD} contains $88,703$ retina images. 
In this work, we only consider images that have diabetic retinopathy, which is a total of $23,359$ images. 
Furthermore, we randomly choose $1,000$ images to train target models.
Note that images in all datasets are resized to $64 \times 64$.

\subsection{Metrics}
\noindent
\textbf{Evaluation metrics for diffusion models.}
We use the popular Fr{\'e}chet Inception Distance~(FID) metric to evaluate the performance of a diffusion model~\cite{FID2017gans}. 
A lower FID score is better, which implies that the generated samples are more realistic and diverse.
Considering the efficiency of sampling, in our work the FID score is computed with all training samples and $1,000$ generated samples.

\smallskip\noindent
\textbf{Evaluation metrics for MI attacks.}
We primarily use the full log-scale receiver operating characteristic~(ROC) curve to evaluate the performance of our attack methods, because it can better characterize the worst-case privacy threats of a victim smodel~\cite{carlini2021membership}.
We also report the true-positive rate~(TPR) at the false-positive rate~(FPR) as it can give a quick evaluation.  
We use average-case metrics --- accuracy as a reference, although it cannot assess the worst-case privacy.

\subsection{Experimental Setup}

In terms of target models, we use open source codes~\cite{sde_code} to train diffusion models, and their recommended hyperparameters about training and sampling are adopted.  
The number of training steps for all models is fixed at $500,000$.
For discrete SDEs, $T$ is fixed as $1,000$ while $T$ is set as 1 for continuous SDEs.
In terms of our attack methods, we evaluate the attack performance using all training samples as member samples and equal numbers of nonmember samples. 
The source code will be made public along with the final version of the paper.

\section{Evaluation}
\label{sec:Evaluation}
In this section, we first present the performance of target models.
Then, we show the performance of our two attacks: loss-based and likelihood-based attacks.

\begin{table}[!h]
	\centering
	\caption{The performance of target models on FFHQ.}
	\label{tab:performance_target}
	\renewcommand{\arraystretch}{1.0}
	\scalebox{0.85}{
		
		\begin{tabular}{c|cccc} 
			\toprule
			Target Models & DDPM  & SMLD  & VPSDE & VESDE  \\ 
			\hline
			FID          & 57.88 & 92.81 & 20.27  & 63.37   \\
			\bottomrule
		\end{tabular}
		
	}
\end{table}

\subsection{Performance of Target Models}
\label{ssec:perf_target_models}
Considering their excellent performance in image generation, we choose DDPM~\cite{ho2020denoising}, SMLD~\cite{song2019generative}, VPSDE~\cite{song2020score} and VESDE~\cite{song2020score} as our target models.
They are trained on the FFHQ dataset containing 1k samples.
Target models with the best FID during the training progress are selected to be attacked.
Table~\ref{tab:performance_target} shows the performance of the target models. 
Figure~\ref{fig:target_gen_imgs} in Appendix shows the qualitative results for these target models.
Overall, all target models can synthesize high-quality images.

\subsection{Performance of Loss-based Attack}
\label{ssec:perf_loss_att_ffhq1k}

We present our attack performance from two aspects: TPRs at fixed FPRs for all diffusion steps and log-scale ROC curves at one diffusion step.
The former aims to provide the holistic performance of our attacks in diffusion models.
In contrast, the latter concentrates on one diffusion step and is able to exhaustively show TPR values at a wide range of FPR values, which is key to assessing the worse-case privacy risks of a model.

\begin{figure*}[!t]
	\centering
	\subfigure[DDPM]{
		\includegraphics[width=0.45\columnwidth]{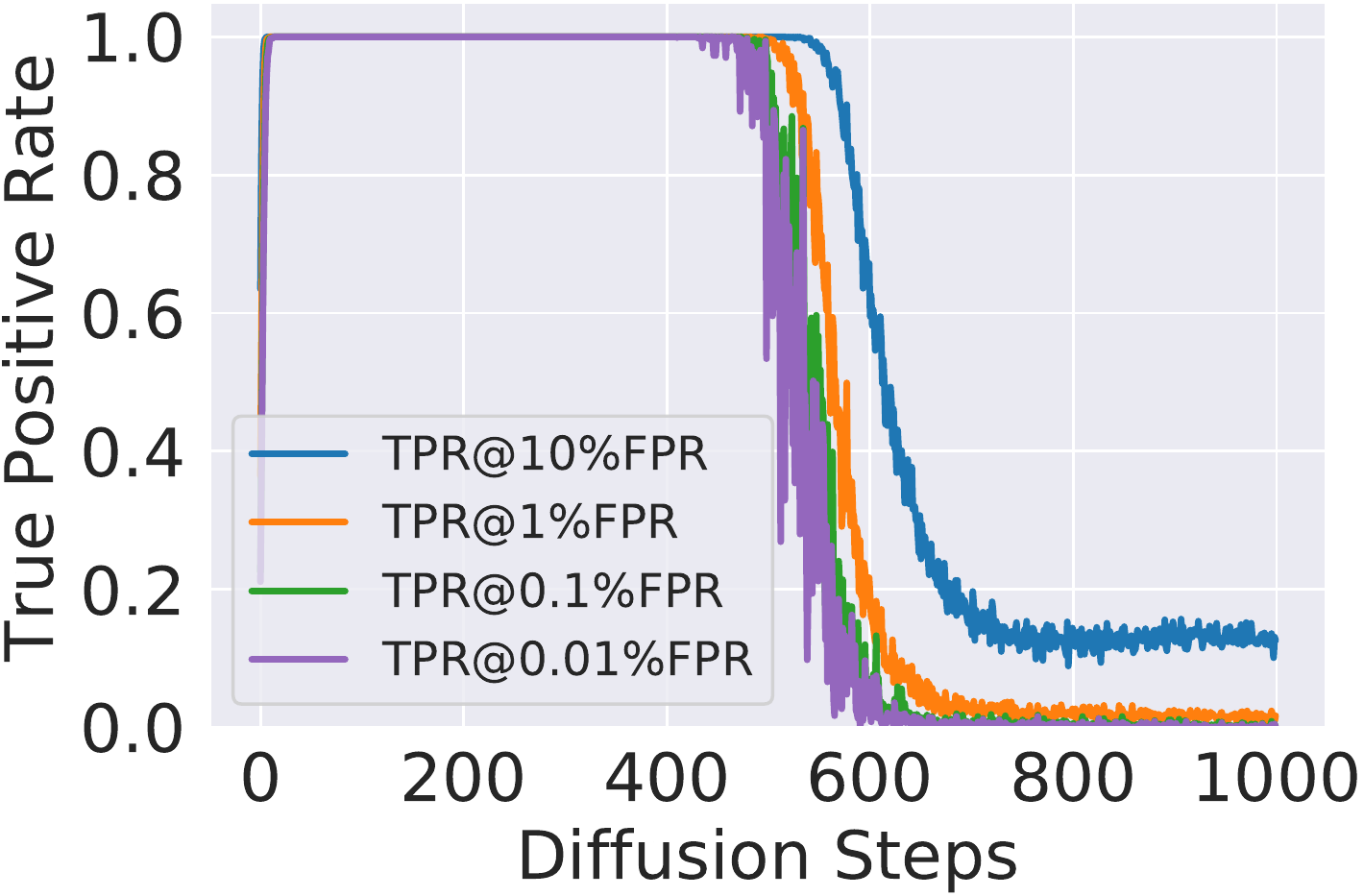}
		\label{fig:loss_ddpm_ffhq1k_TPR}
	}
	\subfigure[SMLD]{
		\includegraphics[width=0.45\columnwidth]{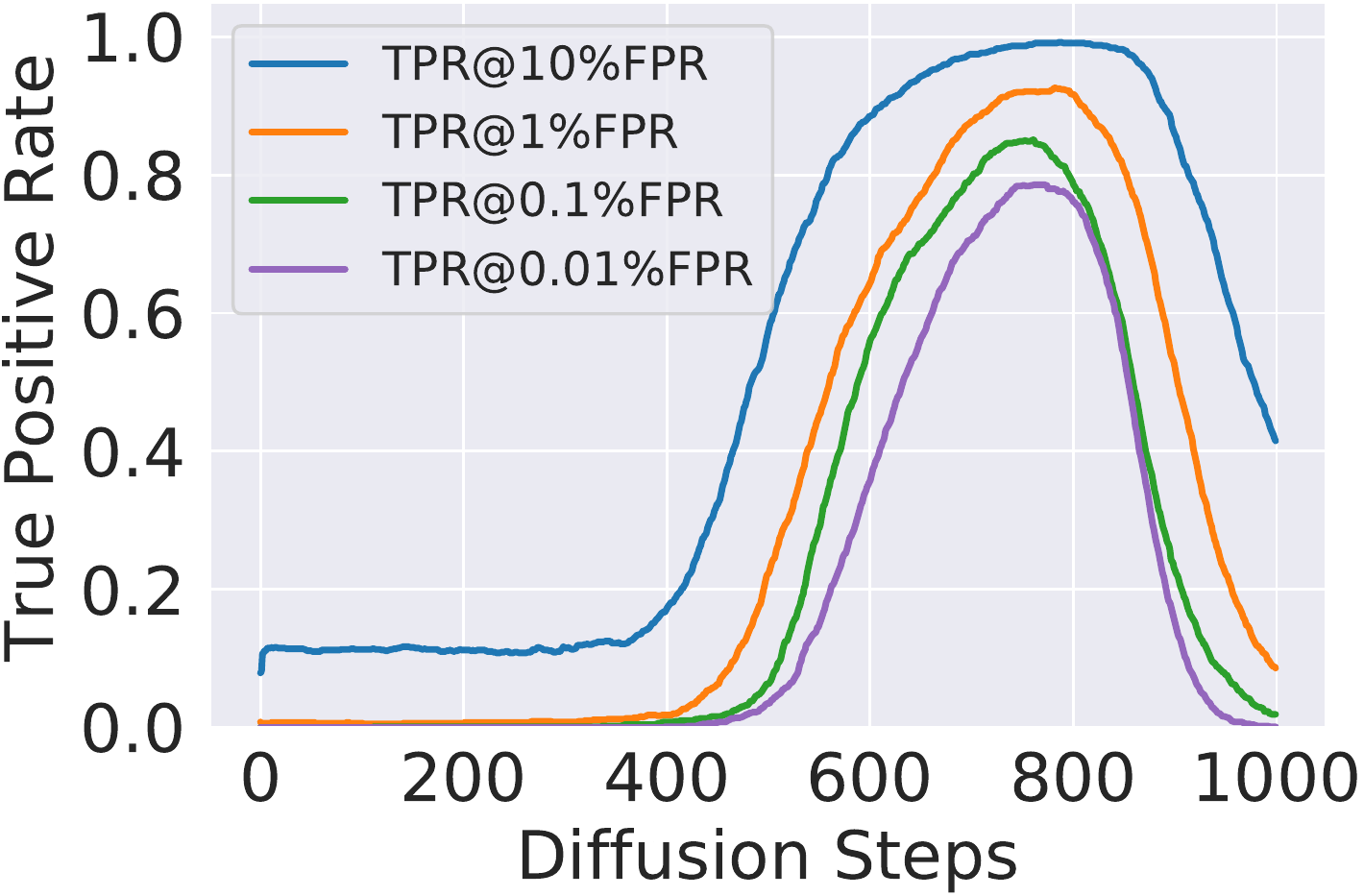}
		\label{fig:loss_smld_ffhq1k_TPR}
	}
	\subfigure[VPSDE]{
		\includegraphics[width=0.45\columnwidth]{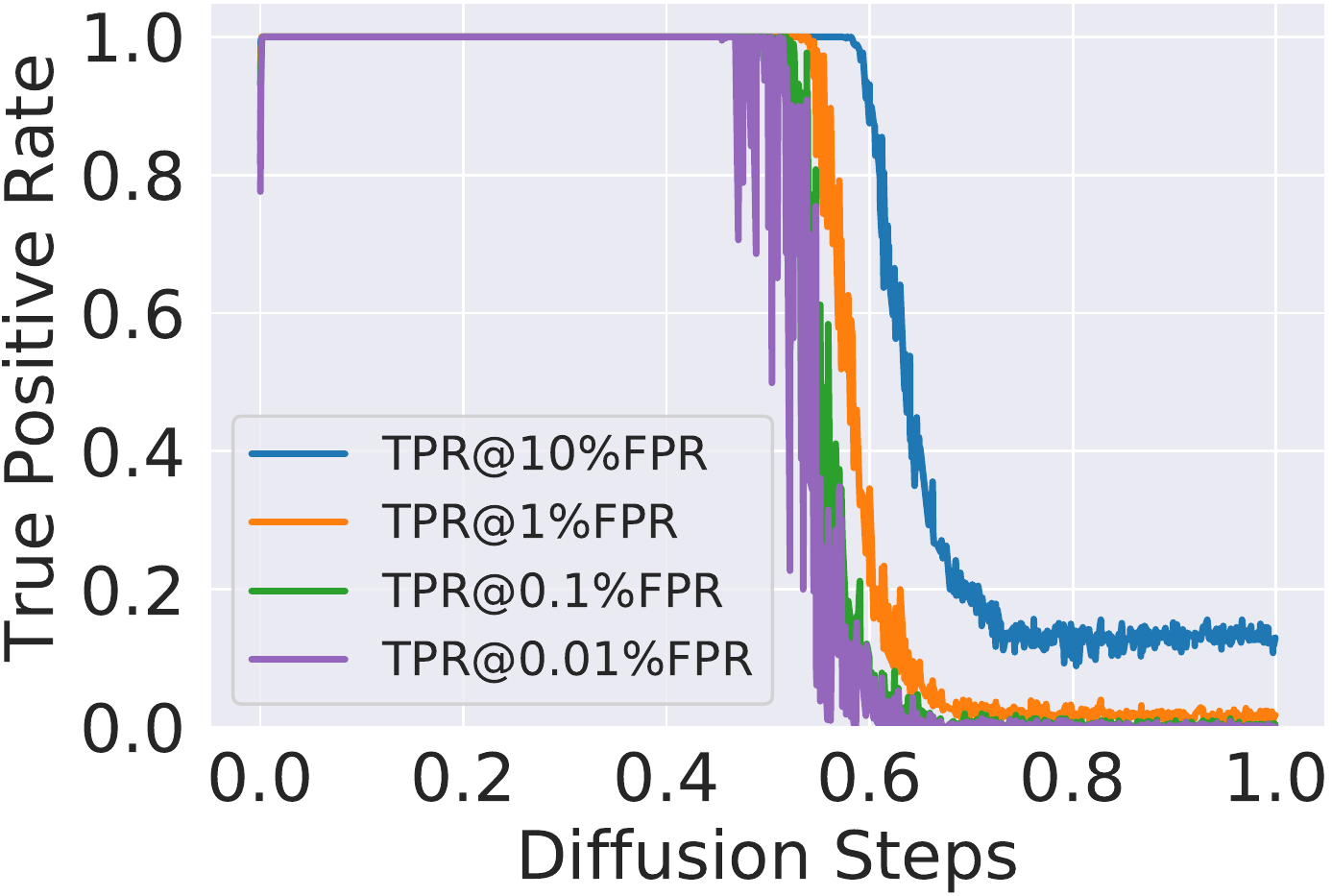}
		\label{fig:loss_vpsde_ffhq1k_TPR}
	}
	\subfigure[VESDE]{
		\includegraphics[width=0.45\columnwidth]{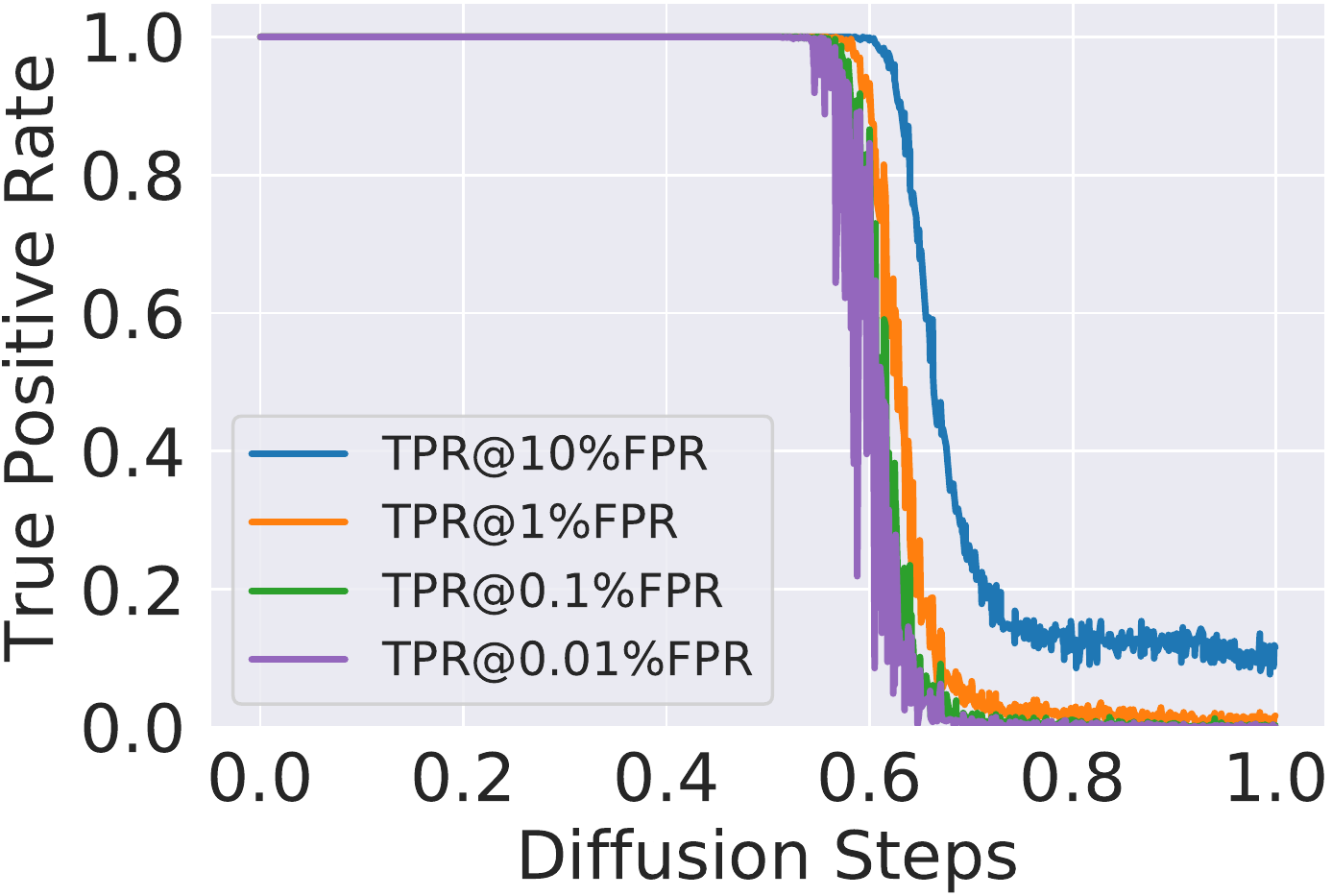}
		\label{fig:loss_vesde_ffhq1k_TPR}
	}
	
	\caption{Performance of the loss-based attack on all diffusion steps. Target models are trained on FFHQ.}
	\label{fig:perf_loss_att}
\end{figure*}

\begin{figure*}[!t]
	\centering
	
	\subfigure[DDPM]{
		\includegraphics[width=0.45\columnwidth]{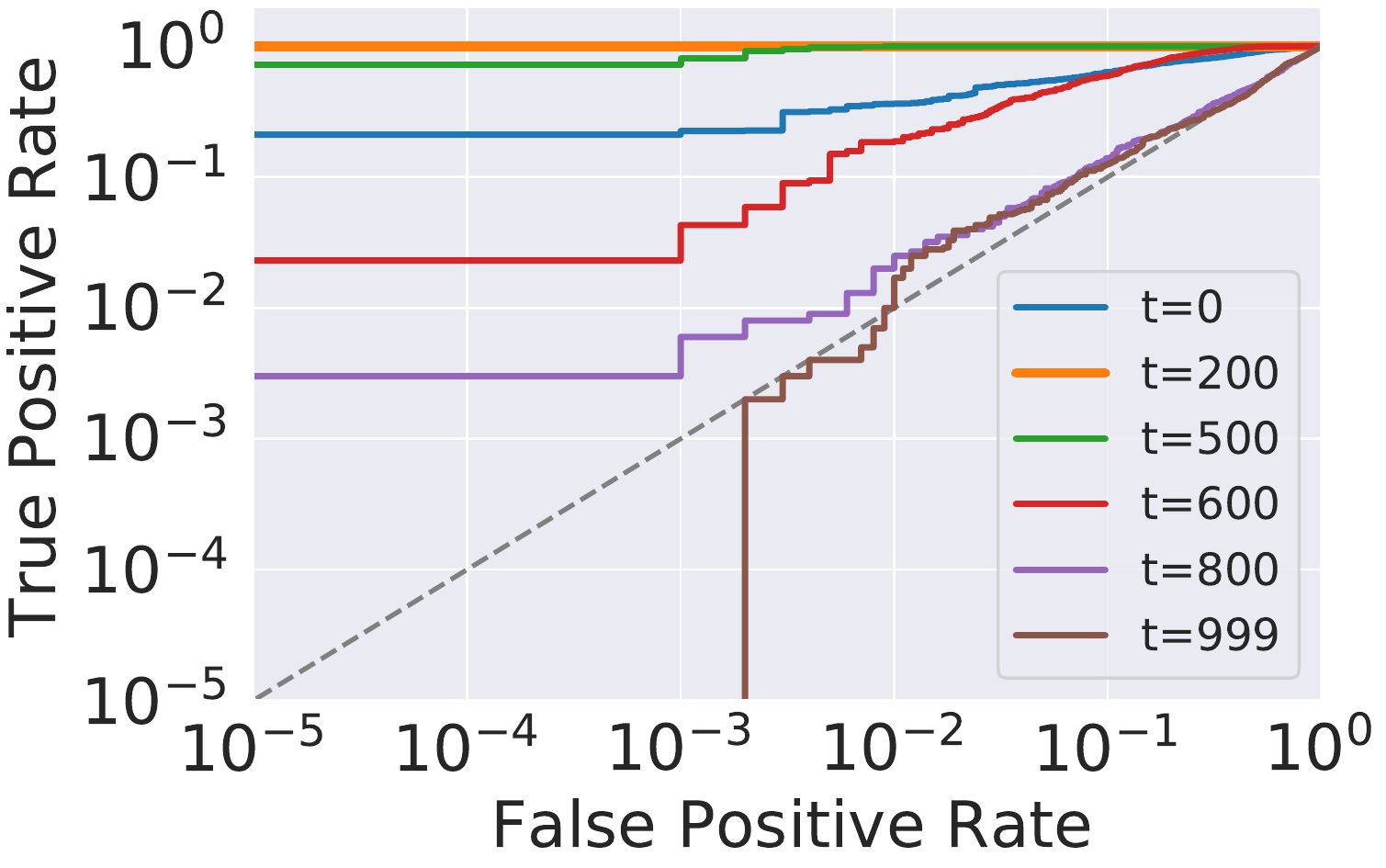}
		\label{fig:loss_ddpm_ffhq1k_TPR_FPR}
	}
	\subfigure[SMLD]{
		\includegraphics[width=0.45\columnwidth]{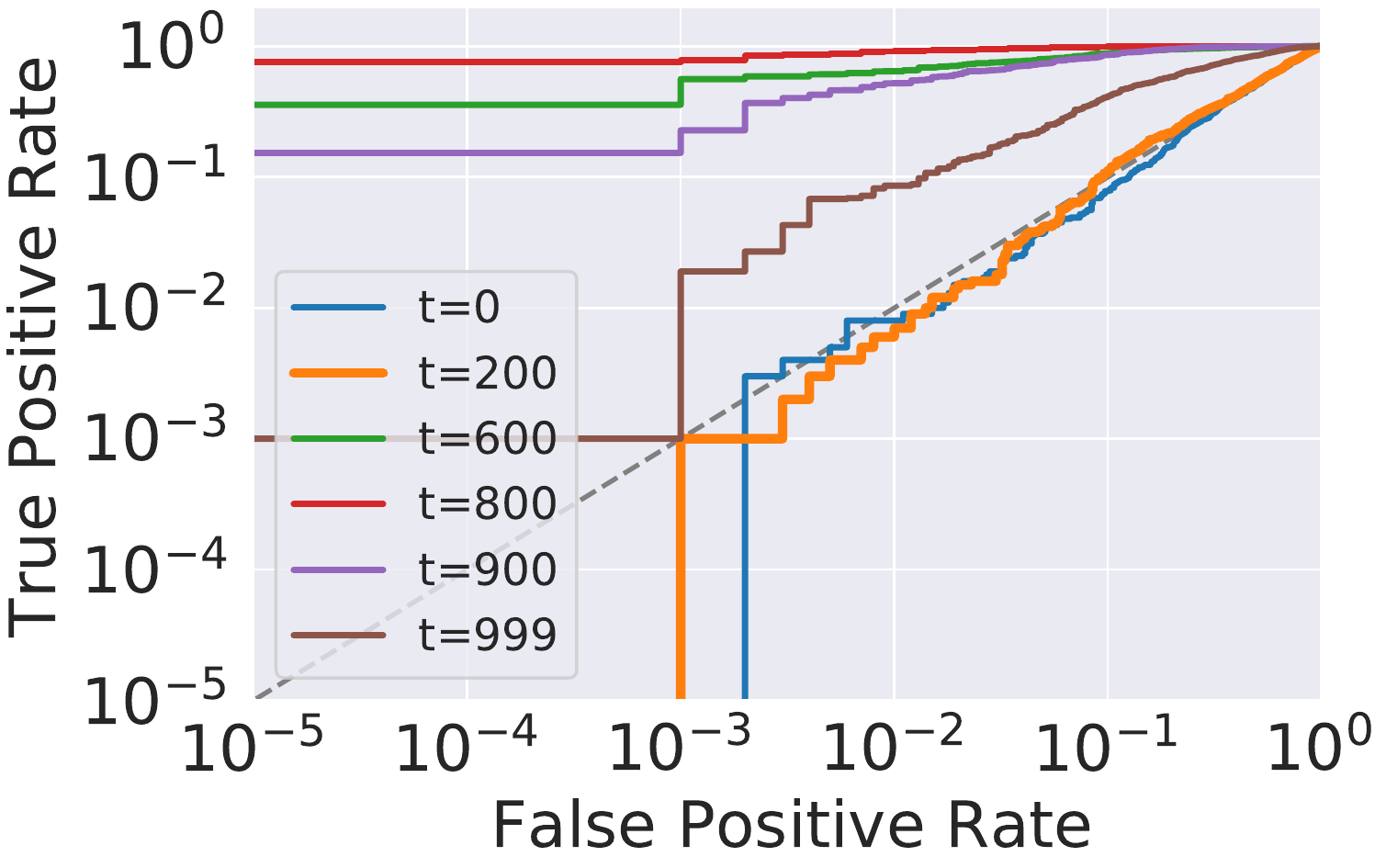}
		\label{fig:loss_smld_ffhq1k_TPR_FPR}
	}
	\subfigure[VPSDE]{
		\includegraphics[width=0.45\columnwidth]{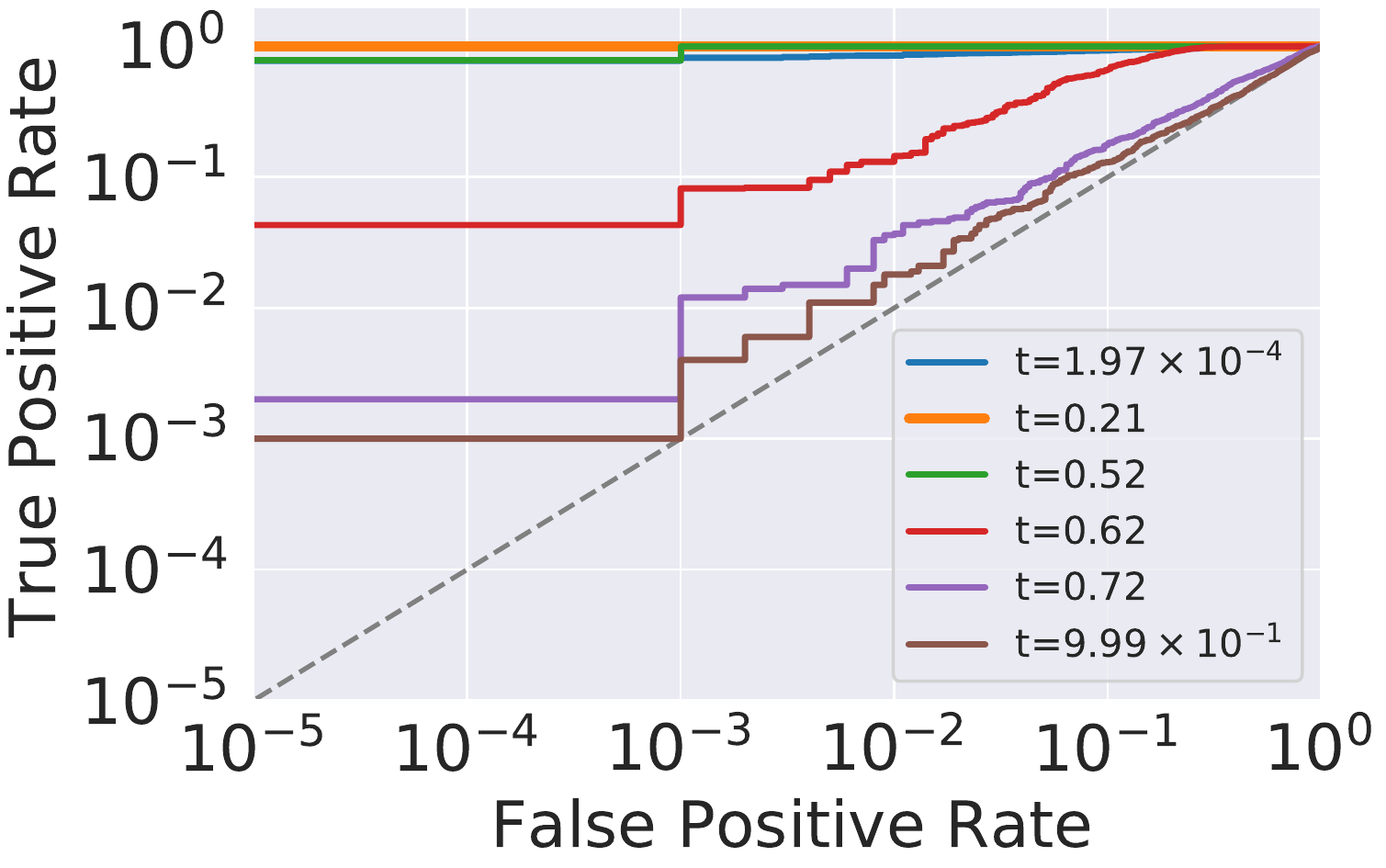}
		\label{fig:loss_vpsde_ffhq1k_TPR_FPR}
	}	
	\subfigure[VESDE]{
		\includegraphics[width=0.45\columnwidth]{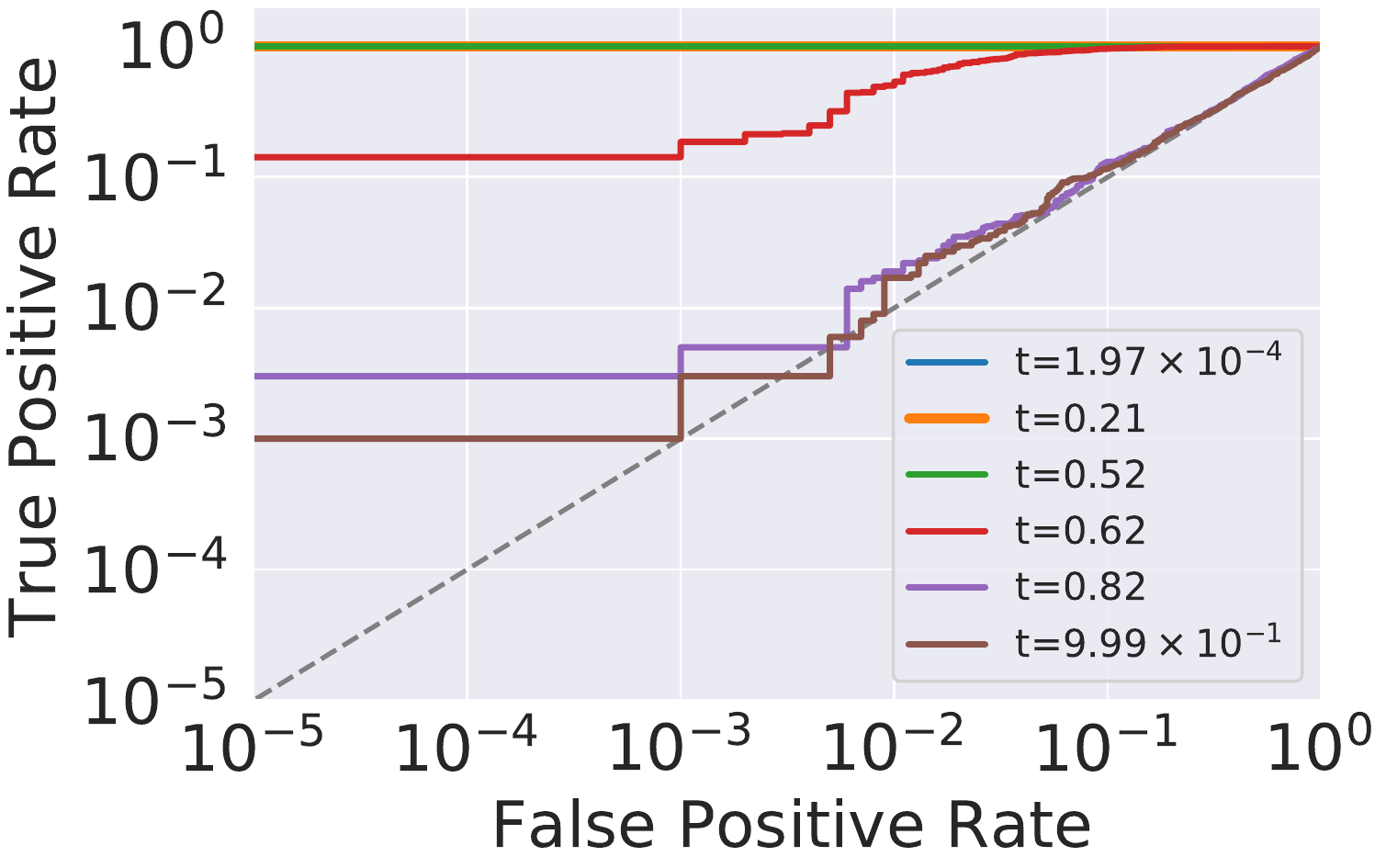}
		\label{fig:loss_vesde_ffhq1k_TPR_FPR}
	}
	
	\caption{Performance of the loss-based attacks at one diffusion step. Target models are trained on FFHQ.}
	\label{fig:perf_loss_att_curve}
\end{figure*}

\noindent
\textbf{TPRs at fixed FPRs for all diffusion steps.} 
Figure~\ref{fig:perf_loss_att} shows the performance of our loss-based attack on four target models trained on FFHQ.
We plot TPRs at different FPRs with regard to diffusion steps for each target model.
Recall DDPM and SMLD models are discrete SDEs  while VPSDE and VESDE models are continuous SDEs.
Thus, the number of diffusion steps for DDPM and SMLD is finite and is fixed as 1,000, while for VPSDE and VESDE models, we uniformly generate $1,000$ points within $[0,1]$ and compute corresponding losses.
Overall, all models are vulnerable to our attacks, even under the worst-case, i.e. TPR at $0.01\%$ FPR, depicted by the purple line of Figure~\ref{fig:perf_loss_att}. 

We observe that our attack presents different performances in different diffusion steps.
There exist high privacy risk regions for diffusion models in terms of diffusion steps.
In these regions~(i.e. diffusion steps), our attack can achieve as high as $100\%$ TPR at $0.01\%$ FPR.
Even for the SMLD model, close to $80\%$ TPR at $0.01\%$ FPR can be achieved.
Recall the training mechanisms of diffusion models. Different levels of noise at different diffusion steps are added during the forward process.
DDPM and VPSDE and VESDE are added growing levels of noise while SMLD starts with maximum levels of noise and gradually decreases the levels of noise.
Thus, we can see that these models (DDPM and VPSDE, and VESDE) 
are more vulnerable to leak training samples in the first half part of the diffusion steps while the SMLD model shows membership vulnerability in the second half part of the diffusion steps.
In brief, all models are prone to suffer from membership leakage in low levels of noise while they become more resistant in high levels of noise. 
In fact, in these diffusion steps where high levels of noise are added to training data, perturbed data is almost close to noise, which can to some degree enhance the privacy of training data.
We also notice that at the starting diffusion step, our attack performance suffers from a decrease. 
This is because there is an instability issue at this step during the training process, which is discussed in the work~\cite{song2020score}.
Despite this, these peak regions still show the effectiveness of our attack.

In addition, as shown in Figure~\ref{fig:perf_loss_att}, we also see that four curves that represent the true positive rate at different false positive rates almost overlap or are very close in most diffusion steps.
It indicates that our attack can still be effective and robust even at the low false positive rate regime.
Note that TPR at low FPR is able to characterize the worst-case privacy risks.

\smallskip\noindent
\textbf{Log-scale ROC curves at one diffusion step.} Figure~\ref{fig:perf_loss_att_curve} plots full log-scale ROC curves of the loss-based attack on four target models.
We choose six different diffusion steps for each target model. 
The rules of choosing diffusion steps for discrete SDEs~(i.e. DDPM and SMLD) are: starting and ending diffusion step and the diffusion step that experiences significant changes in terms of attack performance.
For continuous SDEs~(i.e. VPSDE and VESDE), we first get $1,000$ points that uniformly are sampled from $[0,1]$. Then,
we choose diffusion steps from these points based on the same rule of discrete SDEs.
Overall, our excellent attack performance is exhaustively shown through log-scale ROC curves.

We can observe that when the levels of noise are not too large, our method can achieve a perfect attack, such as at $t=200$ for the DDPM model, $t=800$ for the SMLD model, and $t=0.21$ for the VPSDE and VESDE models.
Again, we can clearly see that the ROC curves on all target models are more aligned with the grey diagonal line with the increase in the magnitudes of noise.
The grey diagonal line means that the attack performance is equivalent to random guesses.
For example, the ROC curves are almost close to the grey diagonal line when the maximal level of noise is added, such as the DDPM model at $t=999$, the SMLD model at $t=0$, and the VPSDE and VESDE models at $t=9.99\times 10^{-1}$.
It is not surprising because at that time the input samples are perturbed as Gaussian noise data in theory and indeed do not have something with original training samples.

We also see that in certain ROC curves, even with the decrease in the FPR values, the TPR values still remain high.
It indicates that the attack is still powerful even in the worst-case. 
Take the VPSDE model at $t=0.72$ as an example, the TPR is $2 \times 10^{-3}$ at the FPR value of $10^{-5}$, which is $20$ times more powerful than random guesses.

\begin{table*}
	\centering
	\caption{Performance of the loss-based attack on target models trained on FFHQ.} 
	\label{tab:TPRatFPR_loss_att}
	\renewcommand{\arraystretch}{1.3}
	\scalebox{0.60}{
		
		\begin{tabular}{lc|rrrrr||lc|rrrrr} 
			\toprule
			Models                 & T                  & TPR@     & TPR@     & TPR@     & TPR@      & Accuracy     & Models                 & T                  & TPR@     & TPR@     & TPR@     & TPR@      & Accuracy      \\
			&                    & 10\%FPR  & 1\%FPR   & 0.1\%FPR & 0.01\%FPR &          &                        &                    & 10\%FPR  & 1\%FPR   & 0.1\%FPR & 0.01\%FPR &           \\ 
			\hline
			\multirow{6}{*}{DDPM}  & 0                  & 63.50\%  & 36.40\%  & 22.50\%  & 21.10\%   & 78.25\%  & \multirow{6}{*}{SMLD}  & 0                  & 7.90\%   & 0.80\%   & 0.00\%   & 0.00\%    & 51.20\%   \\
			& 200                & 100.00\% & 100.00\% & 100.00\% & 100.00\%  & 100.00\% &                        & 200                & 11.20\%  & 0.70\%   & 0.10\%   & 0.00\%    & 52.30\%   \\
			& 500                & 100.00\% & 99.50\%  & 80.80\%  & 72.50\%   & 99.30\%  &                        & 500                & 88.50\%  & 64.40\%  & 56.10\%  & 35.70\%   & 89.50\%   \\
			& 600                & 59.50\%  & 18.80\%  & 4.30\%   & 2.30\%    & 81.15\%  &                        & 800                & 99.10\%  & 91.70\%  & 78.60\%  & 76.10\%   & 96.40\%   \\
			& 800                & 13.90\%  & 2.50\%   & 0.60\%   & 0.30\%    & 52.80\%  &                        & 900                & 85.80\%  & 52.00\%  & 22.80\%  & 15.30\%   & 88.80\%   \\
			& 999                & 12.60\%  & 1.70\%   & 0.00\%   & 0.00\%    & 52.45\%  &                        & 999                & 41.50\%  & 8.60\%   & 1.90\%   & 0.10\%    & 70.55\%   \\ 
			\hline\hline
			\multirow{6}{*}{VPSDE} & $1.97\times 10^{-4}$ & 93.00\%  & 85.00\%  & 81.60\%  & 77.60\%   & 93.15\%  & \multirow{6}{*}{VESDE} & $1.97\times 10^{-4}$ & 100.00\% & 100.00\% & 100.00\% & 100.00\%  & 100.00\%  \\
			& 0.21               & 100.00\% & 100.00\% & 100.00\% & 100.00\%  & 100.00\% &                        & 0.21               & 100.00\% & 100.00\% & 100.00\% & 100.00\%  & 100.00\%  \\
			& 0.52               & 100.00\% & 100.00\% & 99.50\%  & 78.40\%   & 99..90\% &                        & 0.52               & 100.00\% & 100.00\% & 100.00\% & 99.90\%   & 99.95\%   \\
			& 0.62               & 66.50\%  & 14.50\%  & 8.20\%   & 4.30\%    & 85.70\%  &                        & 0.62               & 96.00\%  & 53.60\%  & 18.60\%  & 14.20\%   & 93.25\%   \\
			& 0.72               & 17.90\%  & 3.70\%   & 1.20\%   & 0.20\%    & 57.30\%  &                        & 0.82               & 13.10\%  & 1.90\%   & 0.50\%   & 0.30\%    & 52.50\%   \\
			& $9.99\times 10^{-1}$ & 13.00\%  & 1.8\%    & 0.40\%   & 0.10\%    & 52.20\%  &                        & $9.99\times 10^{-1}$ & 11.60\%  & 1.70\%   & 0.30\%   & 0.10\%    & 51.50\%   \\
			\bottomrule
		\end{tabular}
		
	}
\end{table*}

Table~\ref{tab:TPRatFPR_loss_att} summarizes our attack performance on four target models with regard to diffusion steps and FPR values. 
We also report the average metric accuracy for reference.
Here, we emphasize that only focusing on average metrics cannot assess the worst-case privacy risks. 
For instance, for the DDPM model at $t=800$,  the attack accuracy is $52.80\%$, which indicates the model at this diffusion step almost does not lead to the leakage of training samples, because it is close to $50\%$~(the accuracy of random guesses).
In fact, the TPR is $3 \times 10^{-3}$ at the false positive rate of $10^{-5}$, which is $30$ times more powerful than random guesses.
It means that adversaries can infer confidently member samples under extremely low false positive rates.

\begin{figure*}[]
	\centering
	\subfigure[DDPM]{
		\includegraphics[width=0.45\columnwidth]{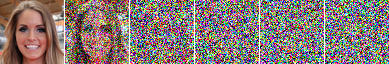}
		\label{fig:perturbed_data_ddpm_ffhq1k}
	}
	\subfigure[SMLD]{
		\includegraphics[width=0.45\columnwidth]{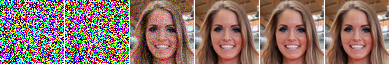}
		\label{fig:perturbed_data_smld_ffhq1k}
	}
	\subfigure[VPSDE]{
	\includegraphics[width=0.45\columnwidth]{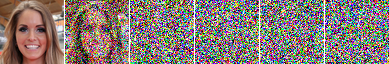}
		\label{fig:perturbed_data_vpsde_ffhq1k}
	}
	\subfigure[VESDE]{
		\includegraphics[width=0.45\columnwidth]{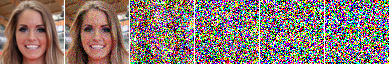}
		\label{fig:perturbed_data_vesde_ffhq1k}
	}
	\caption{Perturbed data of four target models under different diffusion steps.}
	\label{fig:perturbed_data}
\end{figure*}

Figure~\ref{fig:perturbed_data} shows perturbed data of four target models under different diffusion steps.
The diffusion steps in Figure~\ref{fig:perturbed_data} are corresponding to that in Figure~\ref{fig:perf_loss_att_curve}.
We observe that even when some perturbed data that is almost not recognized by human beings is used to train the model, it seems not to prevent model memorization.
For example, for the DDPM model at $t=600$, the perturbed image is meaningless for humans.
However, the attack accuracy is as high as $81.15\%$. At the same time, the TPR at $0.01\%$ FPR is $2.30\%$, which is $230$ times more powerful times than random guesses.
It indicates that models trained on perturbed data, except for Gaussian noise data, can still leak training samples.
The noise mechanism of diffusion models does not provide privacy protection.

\smallskip\noindent
\textbf{Takeaways.} Based on our analysis, when we utilize the loss of a diffusion model to mount a membership inference attack, the loss values from the low levels of noise (about the first $2/3$ of all levels of noise) are a strong membership signal. 
During these diffusion steps, the model shows more privacy vulnerability.

\subsection{Performance of Likelihood-based Attack}

\begin{figure}[!t]
	\centering
	\subfigure[Likelihood-based attack on different target models.]{
		\includegraphics[width=0.45\columnwidth]{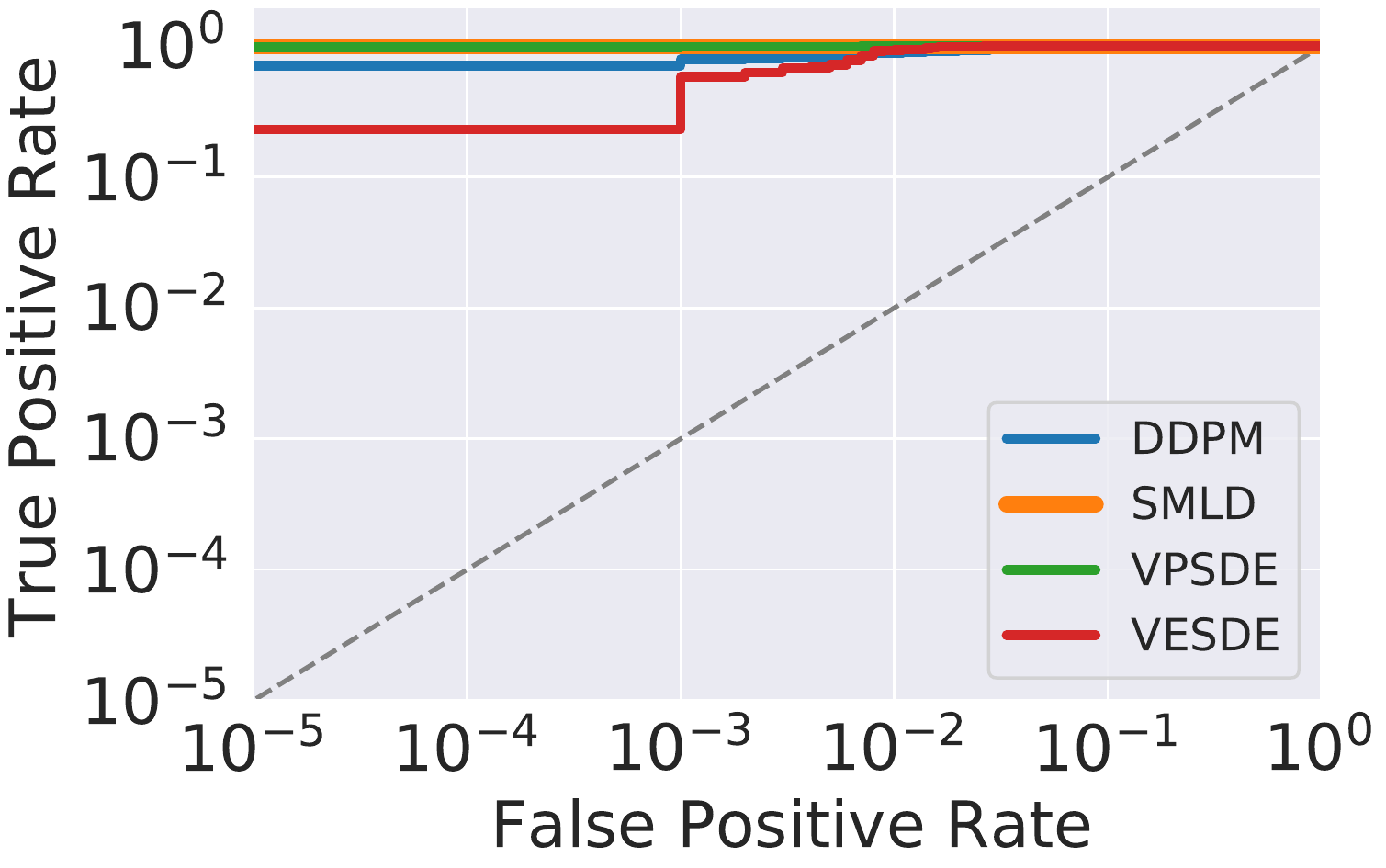}
		\label{fig:likelihood_ffhq1k}
	}
	\subfigure[Likelihood-based attacks on models trained on different sizes of datasets.]{
		\includegraphics[width=0.45\columnwidth]{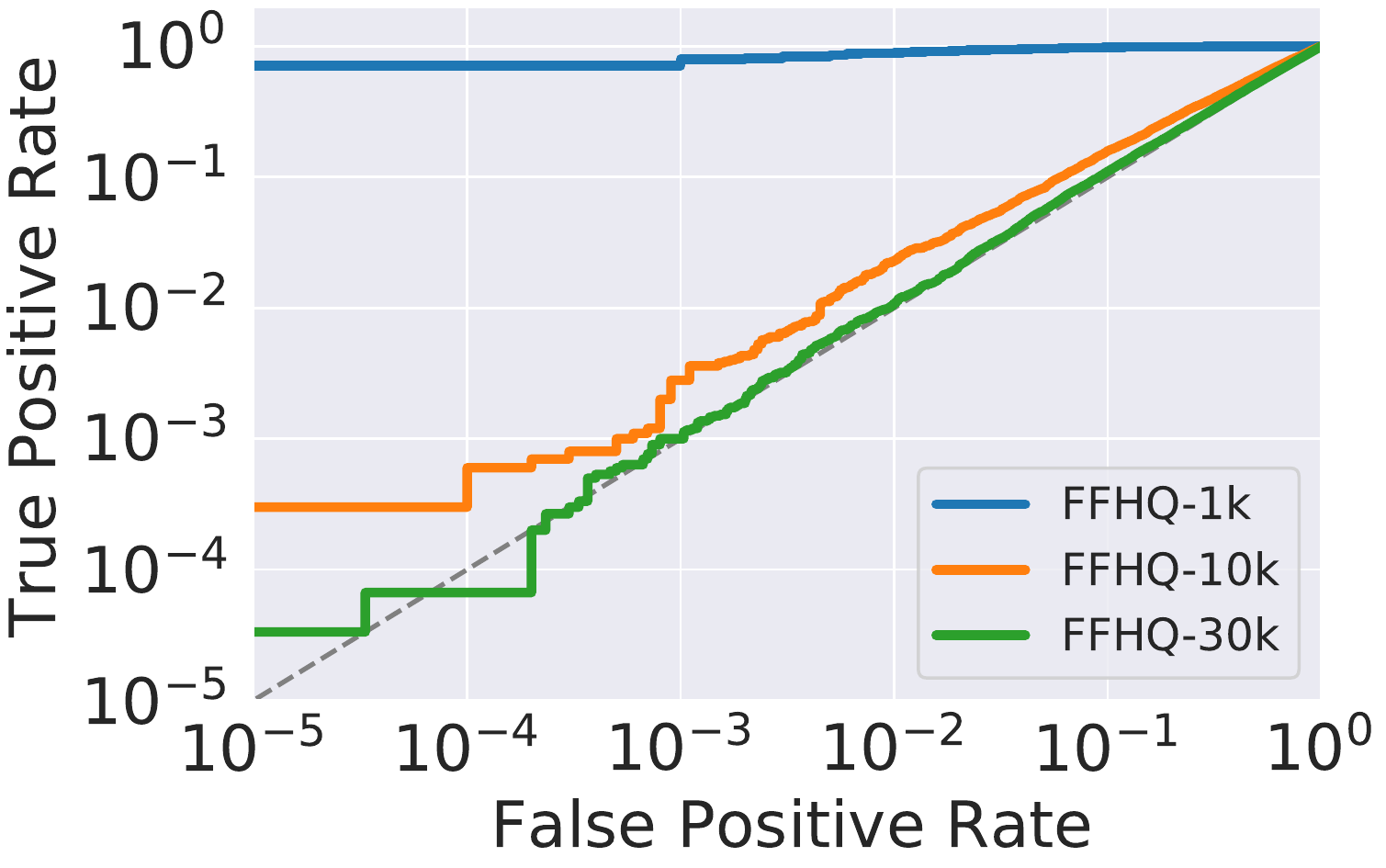}
		\label{fig:likelihood_ffhq_diffsize}
	}
	
	\caption{Performance of the likelihood-based attack.}
	\label{fig:likelihood_ffhq1k_}
\end{figure}

Figure~\ref{fig:likelihood_ffhq1k} demonstrates our likelihood-based attack performance on four target models.
Overall, our attacks still perform well on all target models.
For example, our attack on the SMLD and VPSDE models almost remains $100\%$ true positive rates on all false positive rate regimes.
For the VESDE model, attack results are slightly inferior to the SMLD models, yet still higher than the $10\%$ true positive rate at an extremely low  $0.001\%$ false positive rate.

Table~\ref{tab:likelihood_ffhq1k} shows our attack results at different FPR values for all target models.
Once again, we can clearly see that even at the $0.01\%$ FPR, the lowest TPR among all models is as high as $23.10\%$, which is $2,310$ times than random guesses.
In particular, all training samples that used to train the SMLD model can be inferred at $100\%$ accuracy and $100\%$ TPR at all ranges of FPRs.
In addition, we also observe that the attack accuracy is above $98\%$ for all target models.
Our attack results also remind model providers that they should be careful when using likelihood values.

\begin{table}[!h]
	\centering
	\caption{Likelihood-based attack. Target models are trained on FFHQ.}
	\label{tab:likelihood_ffhq1k}
	\renewcommand{\arraystretch}{1.1}
	\scalebox{0.9}{
		
		\begin{tabular}{l|rrrrr} 
			\toprule
			Models & TPR@     & TPR@     & TPR@     & TPR@      & Accuracy       \\
			& 10\%FPR  & 1\%FPR   & 0.1\%FPR & 0.01\%FPR &           \\ 
			\hline
			DDPM   & 98.00\%  & 89.00\%  & 79.70\%  & 71.00\%   & 95.75\%   \\
			SMLD   & 100.00\% & 100.00\% & 100.00\% & 100.00\%  & 100.00\%  \\
			VPSDE  & 100.00\% & 99.60\%  & 98.90\%  & 98.20\%   & 99.45\%   \\
			VESDE  & 100.00\% & 93.80\%  & 58.40\%  & 23.10\%   & 98.50\%   \\
			\bottomrule
		\end{tabular}

	}
\end{table}

\section{Analysis}
\label{sec:Analysis}
In this section, we first analyze our attack performance with regard to the size of a training set.
Then, we report our results on a medical image dataset DRD. 

\subsection{Effects of Size of a Training Dataset}
\label{ssec:effects_size_data}

We study attack performance with regard to different sizes of the training set of a target model.
Here, we choose the DDPM models trained on FFHQ as target models.
We use FFHQ-1k, FFHQ-10k, and FFHQ-30k to represent different sizes of a dataset, which refer to $1,000$, $10,000$, and $30,000$ training samples in each dataset respectively.
The FID of the target model DDPM trained on FFHQ-1k, FFHQ-10k, and FFHQ-30k are $57.88$, $34.34$, and $24.06$, respectively.
In the following, we present the performance of our both attacks.

\begin{figure*}[!t]
	\centering
	
	\subfigure[TPR at 10\% FPR]{
		\includegraphics[width=0.30\columnwidth]{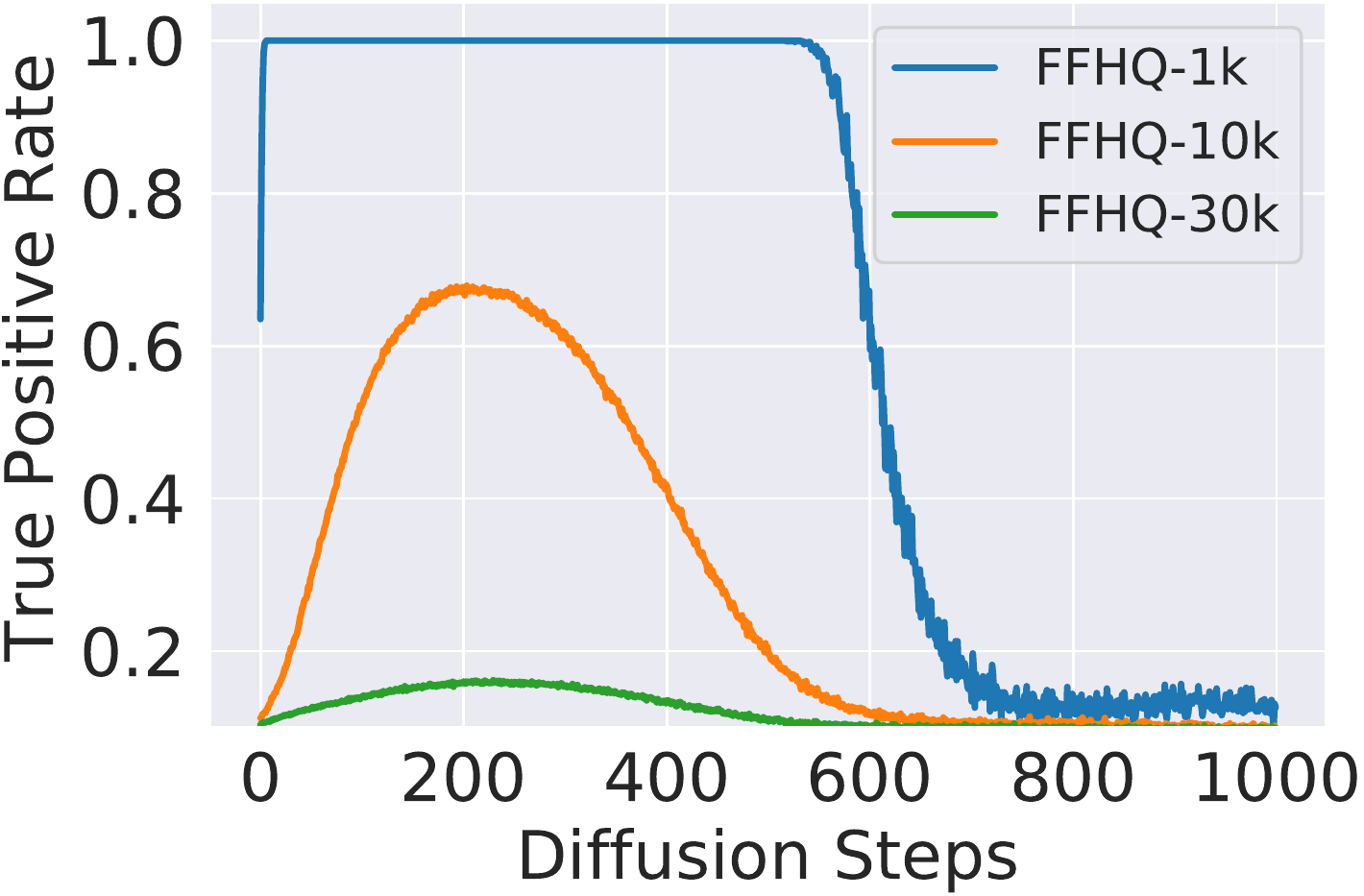}
		\label{fig:loss_ddpm_ffhq_diffsize_TPR_0}
	}	
	\subfigure[TPR at 1\% FPR]{
		\includegraphics[width=0.30\columnwidth]{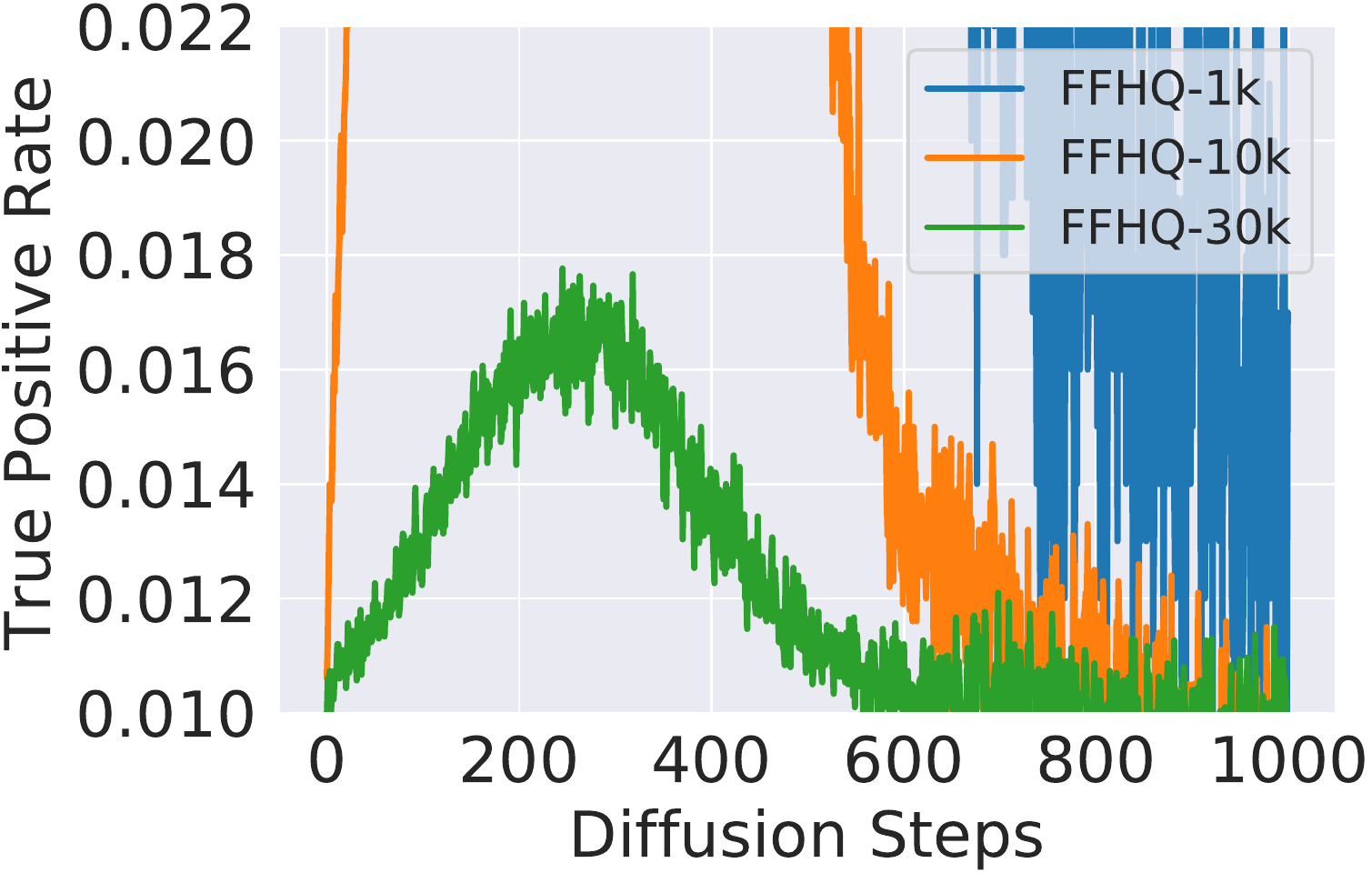}
		\label{fig:loss_ddpm_ffhq_diffsize_TPR_1}
	}
	\subfigure[TPR at 0.1\% FPR]{
		\includegraphics[width=0.30\columnwidth]{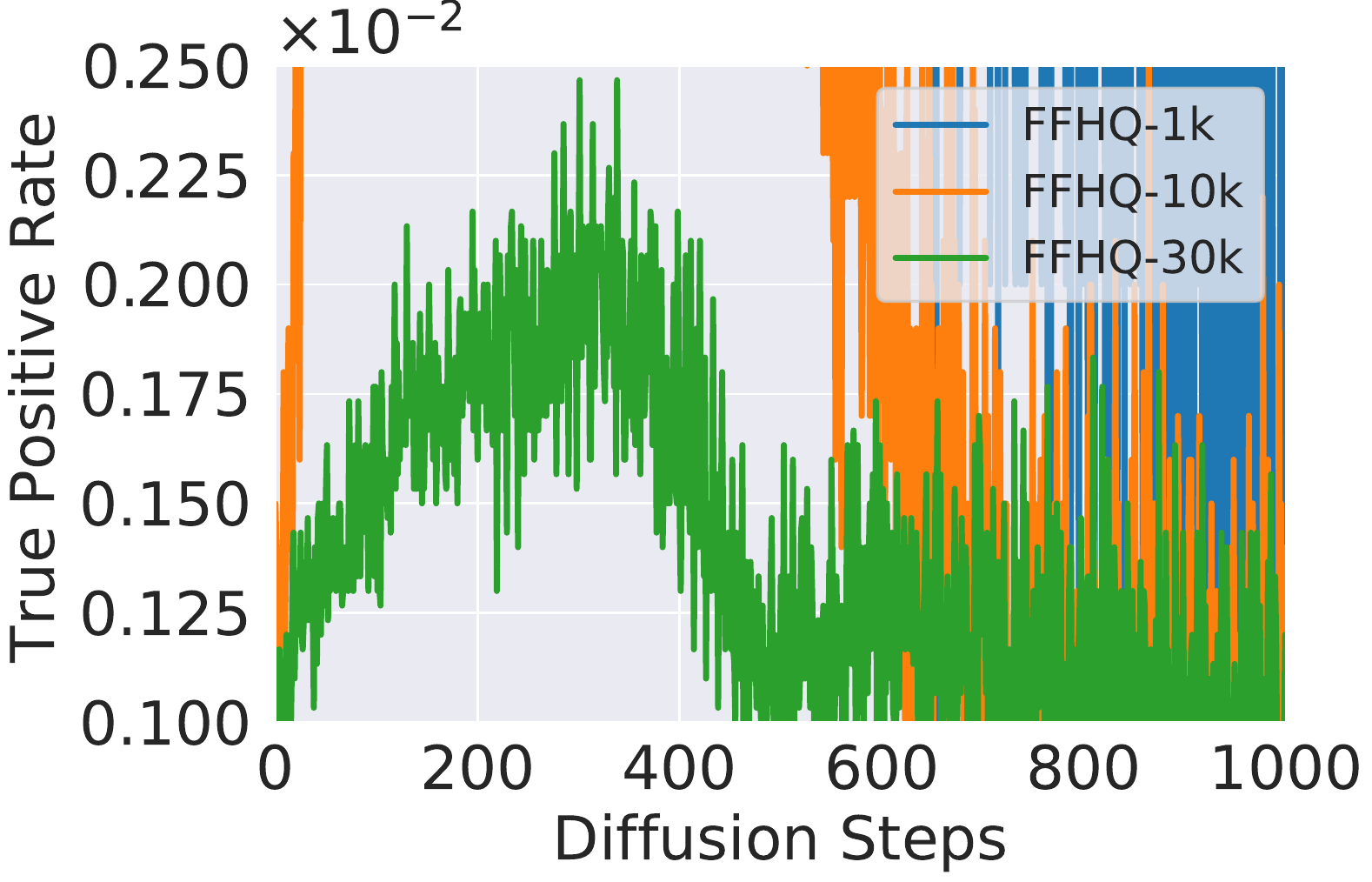}
		\label{fig:loss_ddpm_ffhq_diffsize_TPR_2}
	}	
	\caption{Performance of loss-based attack with different sizes of datasets. The target model is DDPM.}
	\label{fig:loss_ddpm_ffhq_diffsize_TPR}
\end{figure*}

\begin{figure*}[!t]
	\centering
	
	\subfigure[FFHQ-1k]{
		\includegraphics[width=0.30\columnwidth]{fig/loss_ddpm_ffhq1k_TPR_FPR.pdf}
		\label{fig:loss_ddpm_ffhq1k_TPR_FPR_}
	}	
	\subfigure[FFHQ-10k]{
		\includegraphics[width=0.30\columnwidth]{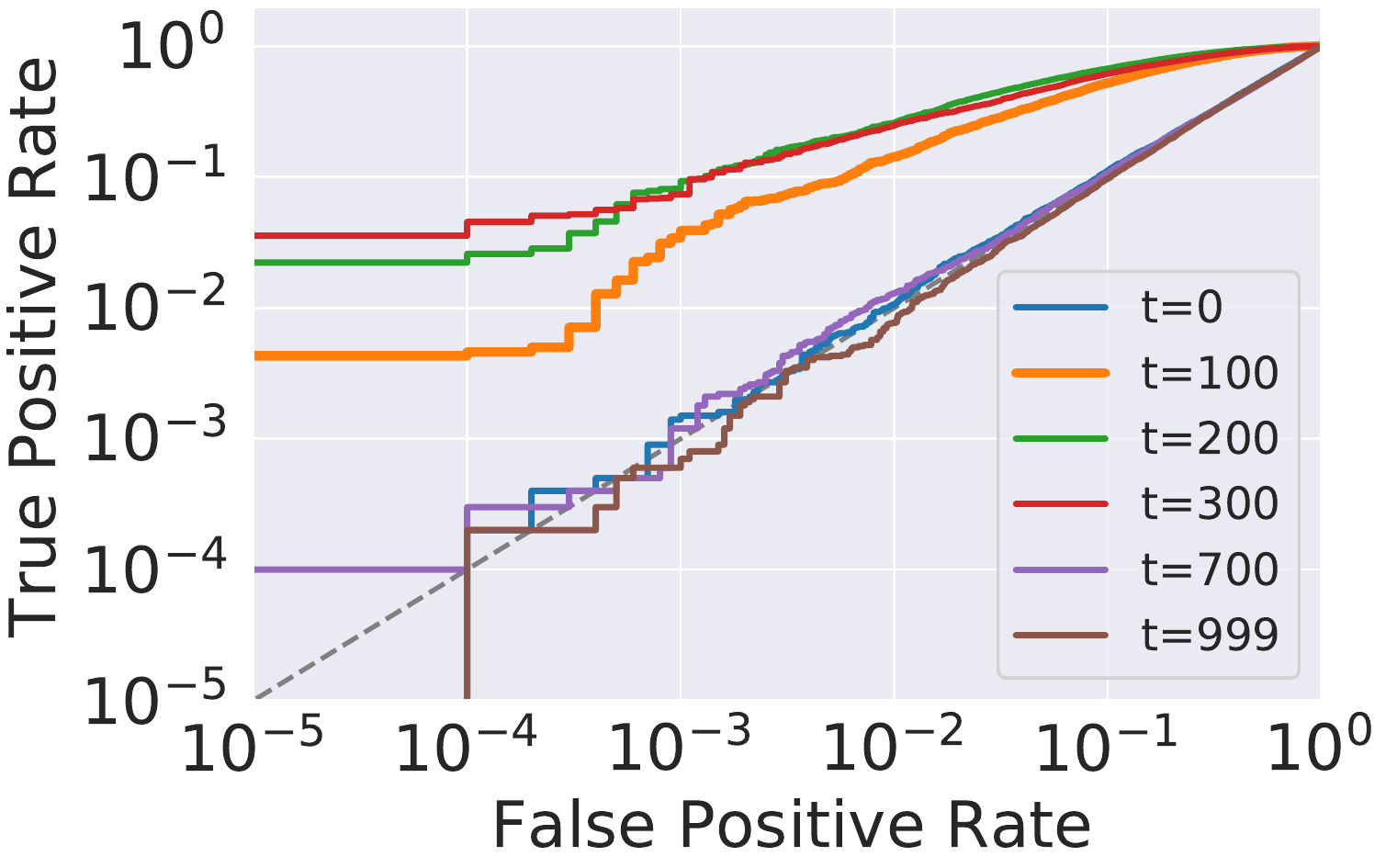}
		\label{fig:loss_ddpm_ffhq10k_TPR_FPR}
	}
	\subfigure[FFHQ-30k]{
		\includegraphics[width=0.30\columnwidth]{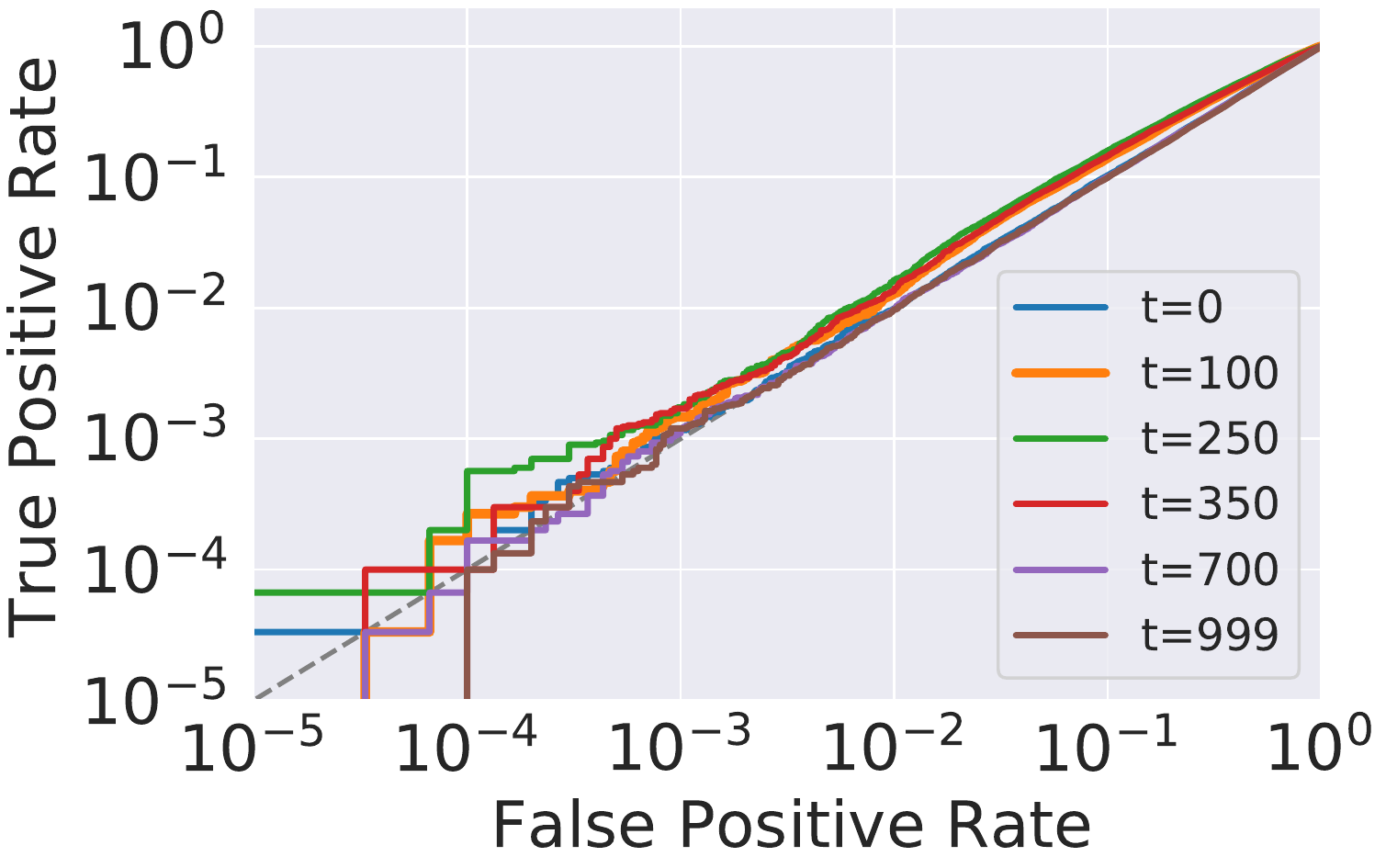}
		\label{fig:loss_ddpm_ffhq30k_TPR_FPR}
	}

	\caption{Performance of loss-based attack with different sizes of datasets. The target model is DDPM. TPR-FPR curves under different time steps.}
	\label{fig:loss_ddpm_ffhq30k_TPR_FPR_diffsize}
\end{figure*}

\smallskip\noindent
\textbf{Performance of loss-based attack.}
Figure~\ref{fig:loss_ddpm_ffhq_diffsize_TPR} depicts the performance of loss-based attacks on all diffusion steps under different sizes of a training set.
Overall, we can observe that attack performance gradually becomes weak when the size of training sets increases.
For example, at diffusion step $t=200$, the TPR at $10\%$ FPR decreases from $100\%$ to about $15\%$ when the training samples increases from $1$k to $30$k.
When the training samples are $10$k, the TPR at $10\%$ FPR still remains above $65\%$.
Similar performances can be seen at $1\%$ FPR and $0.1\%$ FPR.
Here, note that the starting points of the y-axis in Figure~\ref{fig:loss_ddpm_ffhq_diffsize_TPR} are not 0 and we set them as the probability of random guesses.
Thus, the lines that can be shown in the figure indicate this is an effective attack, at least better than random guesses.
We further show the attack performance based on each dataset as a supplement in Figure~\ref{fig:loss_ddpm_ffhq_diffsize} in Appendix.

As illustrated in Section~\ref{ssec:perf_loss_att_ffhq1k}, the peak regions still exist even if the number of training samples increases to 30k.
For instance, as shown in Figure~\ref{fig:loss_ddpm_ffhq_diffsize_TPR_2}, it shows our attack performance of $0.1\%$ FPR on all models. 
Diffusion steps in the range of $0$ to $400$ are still vulnerable to our attack, compared to other steps. 
It indicates that these diffusion steps indeed lead a model to  more easily leak training data.

Figure~\ref{fig:loss_ddpm_ffhq30k_TPR_FPR_diffsize} shows ROC curves of our attack against target models trained on different sizes of training sets.
Based on the same rules described in Section~\ref{ssec:perf_loss_att_ffhq1k}, we select several different diffusion steps and plot their ROC curves.
We can see that indeed models become less vulnerable as the number of training samples increases.
For instance, Figure~\ref{fig:loss_ddpm_ffhq30k_TPR_FPR} shows the DDPM trained on FFHQ-30K is more resistant to membership inference attacks on the full log-scale TPR-FPR curve.
However, when diffusion step $t$ equals $250$, our attack shows higher attack performance than random guesses at the low false positive rate, such as $10^{-4}$.
This is also corresponding to the peak steps in Figure~\ref{fig:loss_ddpm_ffhq_diffsize_TPR}.

\smallskip\noindent
\textbf{Performance of likelihood-based attack.}
Figure~\ref{fig:likelihood_ffhq_diffsize} shows the performance of likelihood-based attacks in terms of different sizes of training sets.
Similar to the loss-based attack, the performance of the likelihood-based attack decrease with an increase in the sizes of training sets.
Specifically, the likelihood-based attack shows perfect performance on the target model trained on FFHQ-1k.
When the size of a training set increases to $10$K, there is a significant drop but still better than random guesses on the full log-scale ROC curve.
In particular, in the extremely low false positive rate regime, such as $10^{-4}$, the true positive rate is about $6\times 10^{-4}$, which is $6$ times more powerful than random guesses. 
In the model trained on FFHQ-30K, the ROC curve is almost close to the diagonal line, which indicates that adversaries are difficult to infer member samples through likelihood values.

\subsection{Effects of Different Datasets}
\label{ssec:perf_drd}
In this subsection, we show our attack performance on a medical image dataset about diabetic retinopathy.
We have described this dataset DRD in Section~\ref{ssec:Datasets}.
We choose the SMLD as the target model and the number of training samples is $1,000$.
Overall, the SMLD model can achieve excellent performance in image synthesis, with an FID of $33.20$.
Figure~\ref{fig:generated_smld_drd1k} in Appendix visualizes synthetic samples, which all show good quality.

\begin{figure*}[!t]
	\centering
	
	\subfigure[Loss-based attack]{
		\includegraphics[width=0.30\columnwidth]{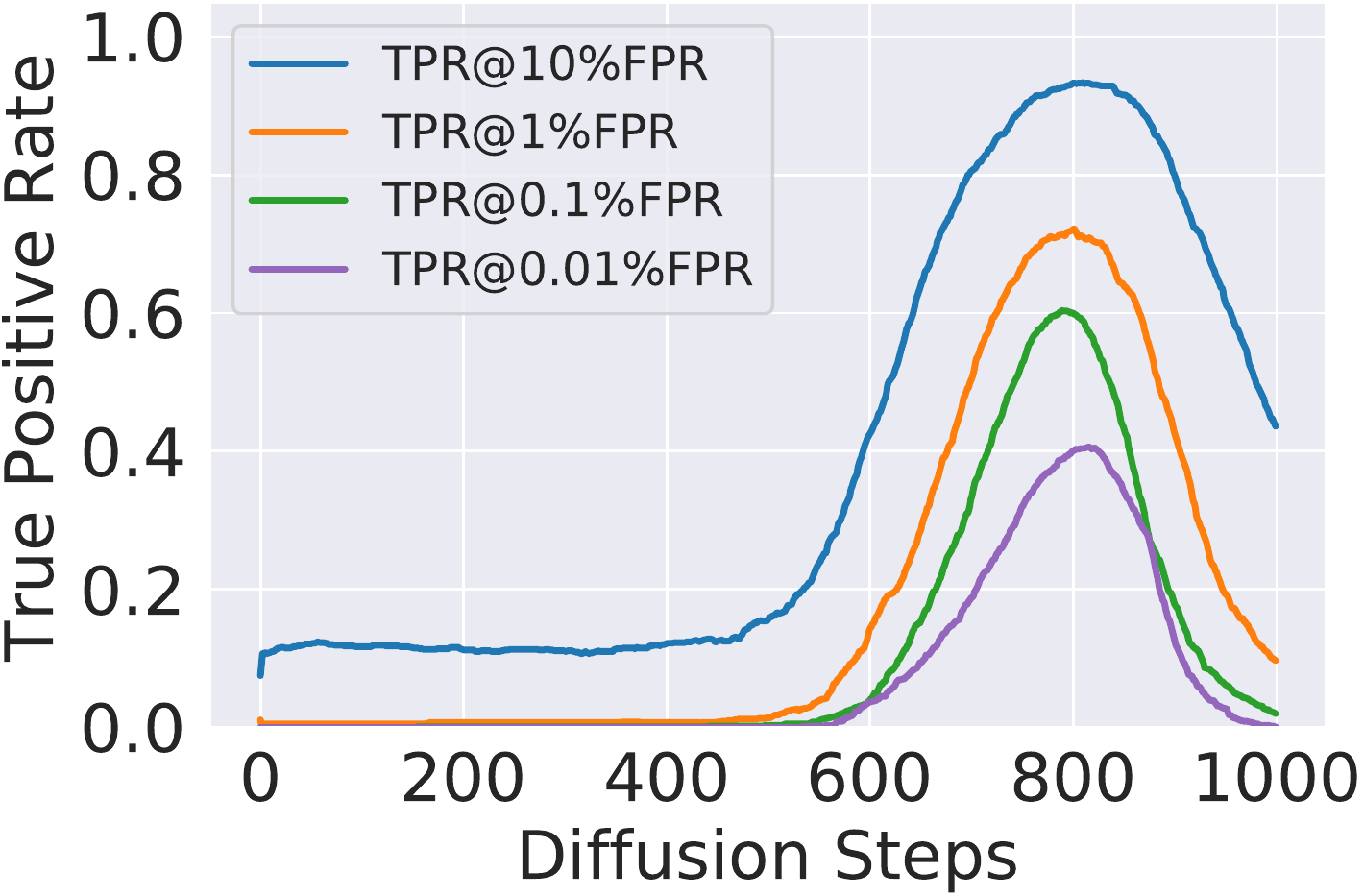}
		\label{fig:loss_smld_drd1k_TPR}
	}	
	\subfigure[Loss-based attack]{
		\includegraphics[width=0.30\columnwidth]{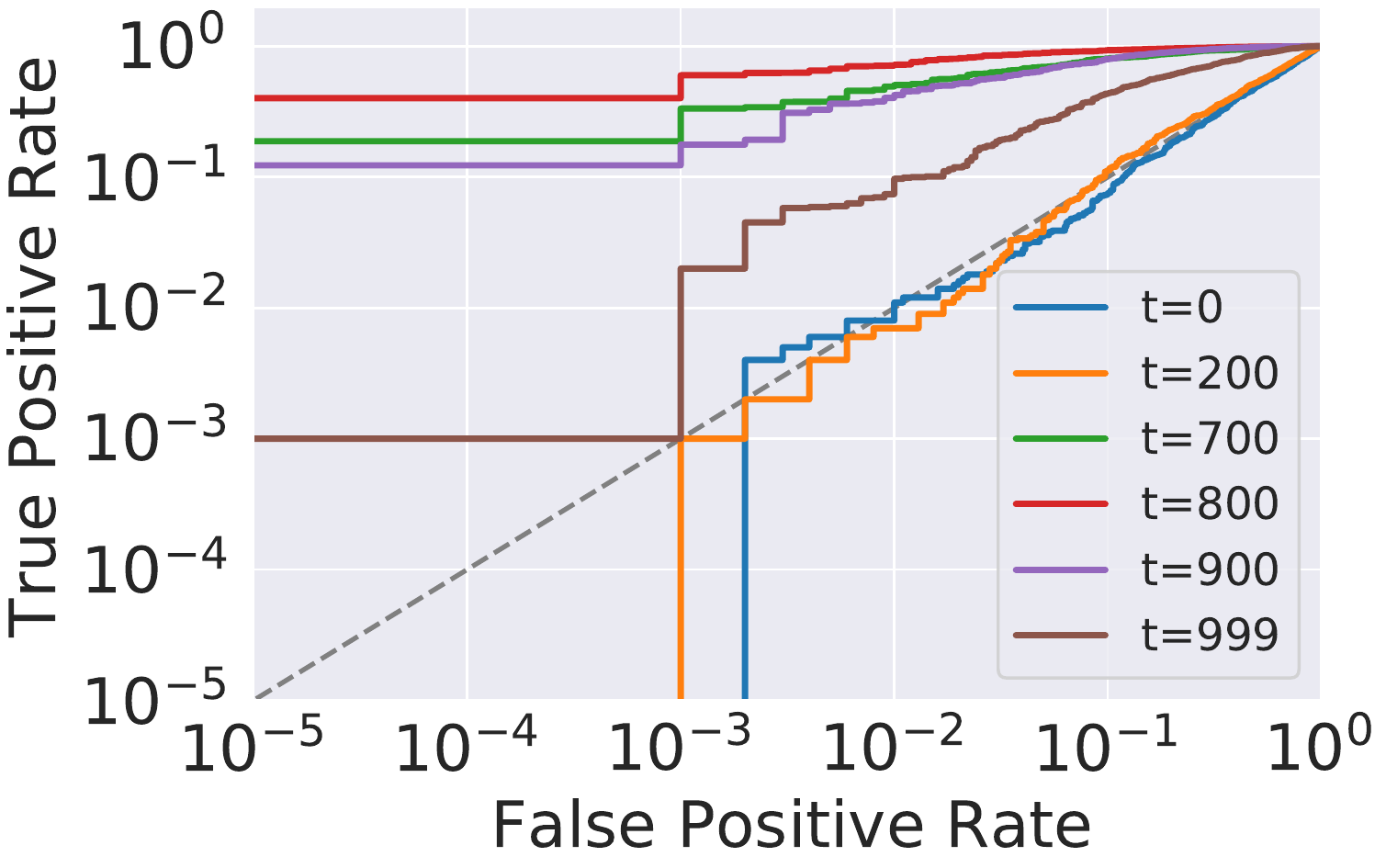}
		\label{fig:loss_smld_drd1k_TPR_FPR}
	}
	\subfigure[Likelihood-based attack]{
	\includegraphics[width=0.30\columnwidth]{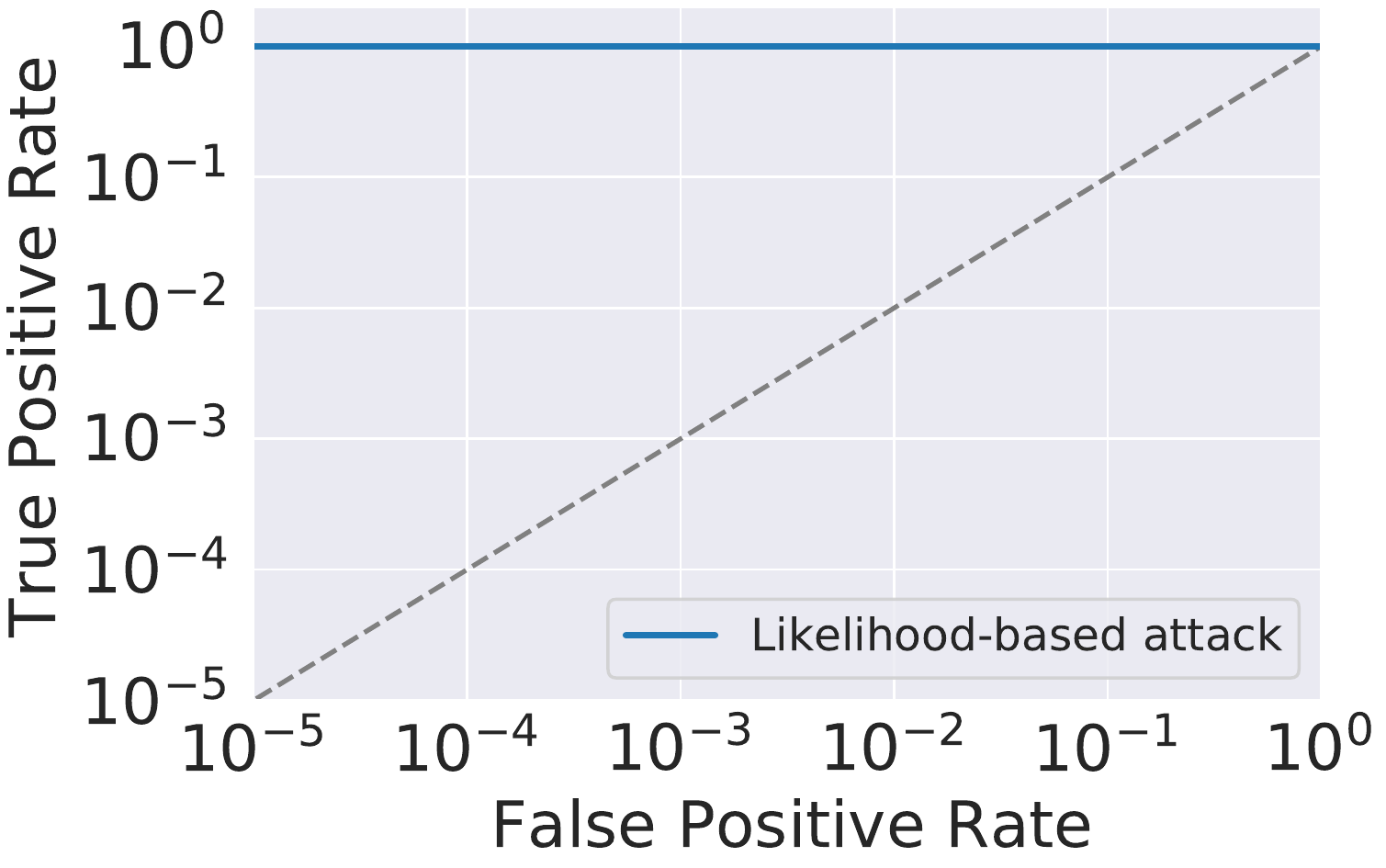}
	\label{fig:likelihood_smld_drd1k}
}
	\caption{Attack performance on DRD.}
	\label{fig:loss_smld_drd1k_diffdataset}
\end{figure*}

\smallskip\noindent
\textbf{Performance of loss-based attack.}
Figure~\ref{fig:loss_smld_drd1k_diffdataset} shows the performance of loss-based attacks for the target model SMLD trained on DRD.
Here, note that the levels of the noise of the SMLD model gradually become small with an increase in diffusion steps.

Figure~\ref{fig:loss_smld_drd1k_TPR} shows the performance of our loss-based attack on all diffusion steps.
We can again observe that our attacks can still perform perfectly on the DRD dataset at diffusion steps of low levels of noise.
For instance, our attack can achieve at least $40\%$ TPR for four different FPRs when diffusion step $t$ is around $800$.
In addition, TPR curves at different FPRs all show the same trend on all diffusion steps.
To be specific, as shown in Figure~\ref{fig:loss_smld_drd1k_TPR}, the true positive rate starts to increase from $t=500$ and reaches a peak at $t=800$. After that, it gradually decreases.

Figure~\ref{fig:loss_smld_drd1k_TPR_FPR} depicts ROC curves for different diffusion steps on target model SMLD trained on DRD.
We can see that for diffusion steps where low levels of noise are added, such as $t=700$, $800$, or $900$, training samples can be inferred at least $10\%$ TPR at as low as $0.001\%$ FPR.
At high levels of noise diffusion step, such as $t=0$, $200$, the performance of our method is almost close to random guesses.

\smallskip\noindent
\textbf{Performance of likelihood-based attack.}
Figure~\ref{fig:likelihood_smld_drd1k} reports the performance of our likelihood-based attack on the SMLD model trained on DRD.
As expected, our attack still shows excellent performance.
We can clearly find that the attack achieves $100\%$ TPR on all FPR values, which means that all member samples are inferred correctly.
Table~\ref{tab:loss_ddpm_drd1k_diffdataset} in Appendix reports the quantitative results of both attacks.

\section{Defenses}
\label{sec:Defenses}

Differential privacy (DP)~\cite{abadi2016deep,dwork2008differential} is regarded as the gold standard for protecting the training set of a model. 
In practice, although differential privacy can guarantee individual-level privacy, it often sacrifices significantly model utility, especially for the quality of generated images, when it is applied to generative models.
In this section, we present our attack results on diffusion models using the DP defense technology. 

We adopt Differentially-Private Stochastic Gradient Descent (DP-SGD)~\cite{abadi2016deep} to train diffusion models.
DP-SGD is widely used for privately training a machine learning model.
Generally, DP-SGD achieves differential privacy by adding noise into per-sample gradients.
In our work, we implement DP diffusion models through the Opacus library~\cite{Yousefpour2021OpacusUD}.
We set the clip bound~$C$ and the failure probability $\delta$ as 1 and $5\times 10^{-4}$. 
The batch size and the number of epochs are $64$ and $1,800$. 
The final privacy budget $\epsilon$ is $19.62$.
We choose the DDPM model as the target model.
It is trained on FFHQ containing $1,000$ training samples, and the FID is $393.94$.

\begin{figure*}[!t]
	\centering
	
	\subfigure[Loss-based attack]{
		\includegraphics[width=0.30\columnwidth]{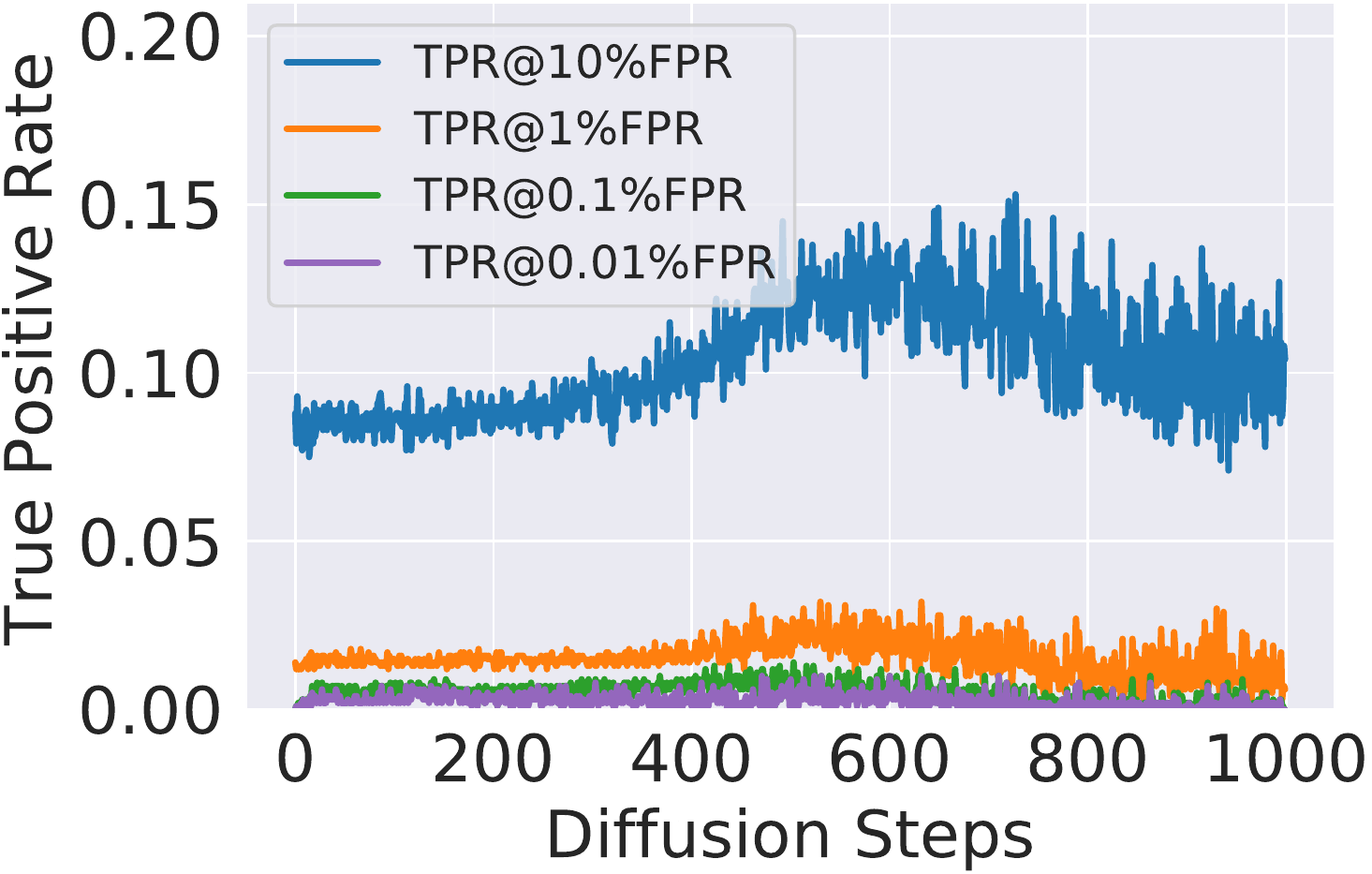}
		\label{fig:dp_loss_ddpm_ffhq1k_TPR}
	}	
	\subfigure[Loss-based attack]{
		\includegraphics[width=0.30\columnwidth]{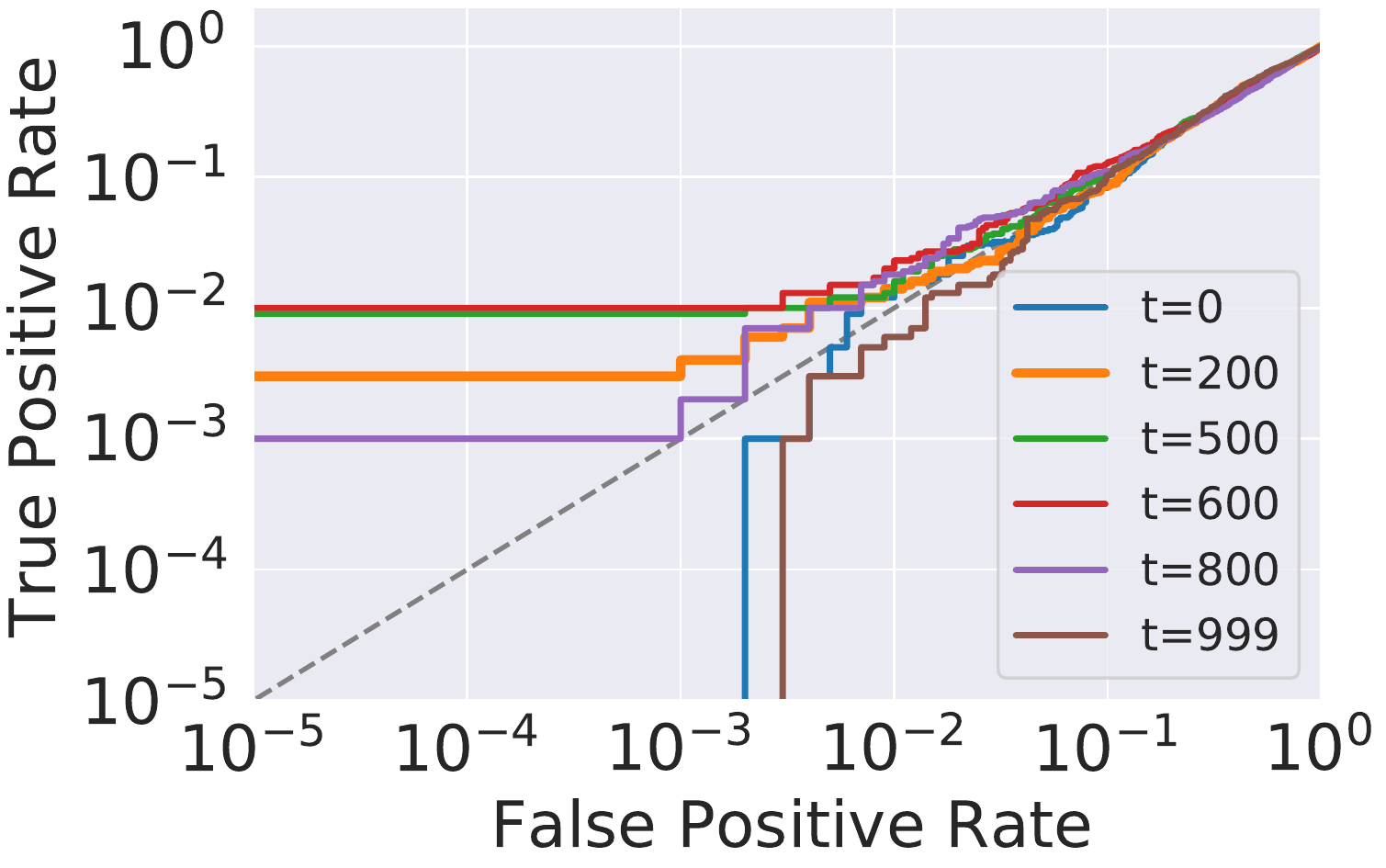}
		\label{fig:dp_loss_ddpm_ffhq1k_TPR_FPR}
	}
	\subfigure[likelihood-based attack]{
		\includegraphics[width=0.30\columnwidth]{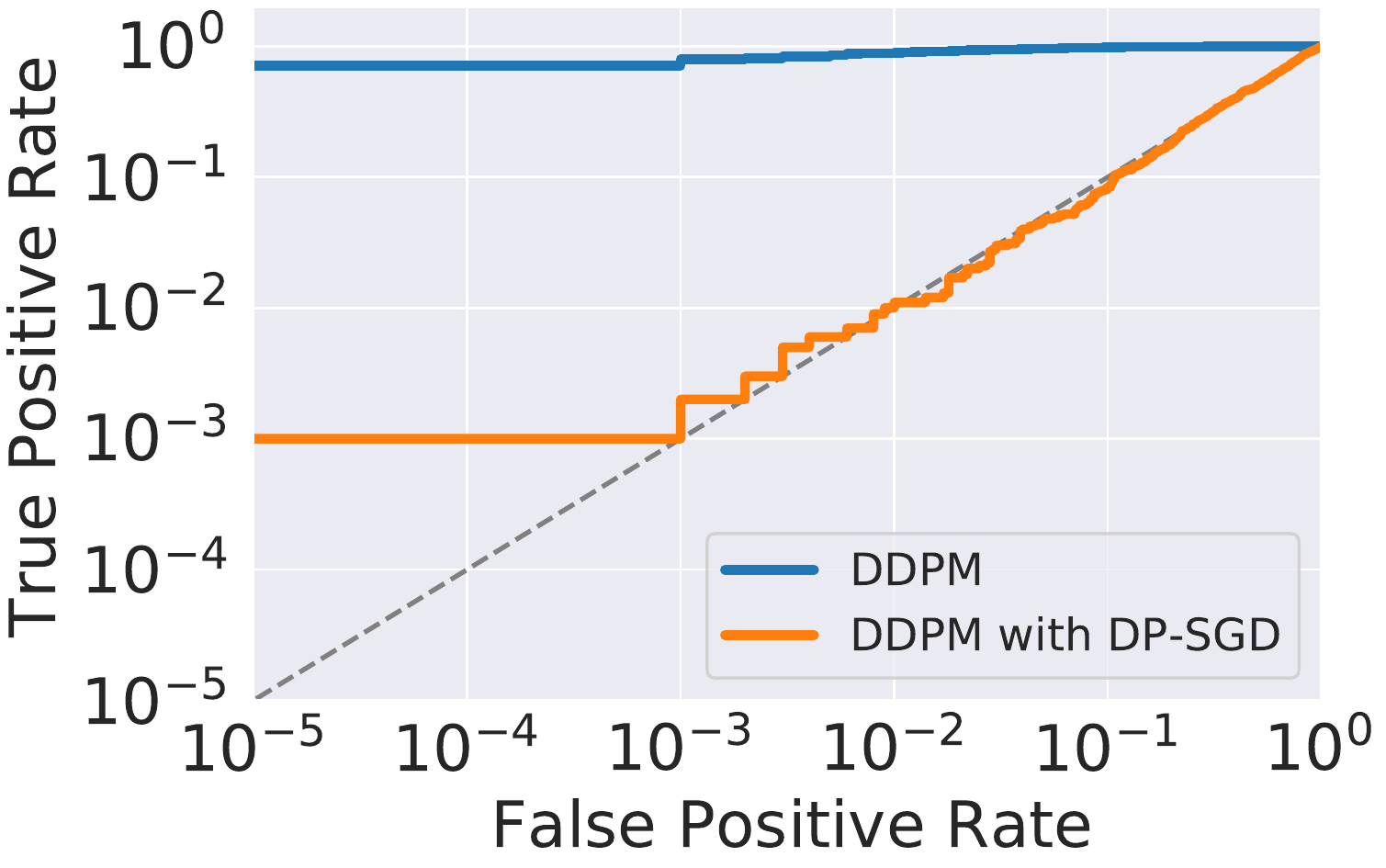}
		\label{fig:dp_likelihood_ffhq1k}
	}		
	\caption{Attack performance on DDPM with DP-SGD.}
	\label{fig:dp_loss_ddpm_ffhq1k}
\end{figure*}

\smallskip\noindent
\textbf{Performance of loss-based attack.}
Figure~\ref{fig:dp_loss_ddpm_ffhq1k} show the performance of both types of attacks on DDPM trained with DP-SGD on FFHQ.
As described in Figure~\ref{fig:dp_loss_ddpm_ffhq1k_TPR}, we present the performance of loss-based attack on all diffusion steps.
Clearly, we can see that differentially training DDPM, i.e. DDPM with DP-SGD indeed can significantly decrease the membership leakages.
The peak regions can be still seen when diffusion steps are between $400$ and $800$.
However, the true positive rates reduce to at most $15\%$ during these diffusion steps when the false positive rate is $10\%$.

Figure~\ref{fig:dp_loss_ddpm_ffhq1k_TPR_FPR} further shows ROC curves of our loss-based attack on different diffusion steps.
Overall, ROC curves still remain close to the diagonal line before the false positive rate is $1\%$, i.e. $10^{-2}$.
It indicates that adversaries almost make a random guess.
When the false positive rates continue to decrease from $0.1\%$ to $0.001\%$, different diffusion steps show divergences.
For diffusion steps at 500 or 600, the true positive rates keep at $1\%$.
In contrast, the true positive rates reduce to $0\%$ at diffusion step $t=999$.
It means that in the worst-case, some training samples are still inferred with a probability higher than random guesses.

\smallskip\noindent
\textbf{Performance of likelihood-based attack.}
Figure~\ref{fig:dp_likelihood_ffhq1k} show the performance of likelihood-based attack on DDPM training with DP-SGD on FFHQ.
Again, we can see that differentially private training a diffusion model indeed can mitigate our attack.
At the same time, we also see at the low false positive rate regime, our attack still remains at 0.1\% true positive rate, which illustrates the effectiveness of our attack even in the worst-case.
Here, we also note that the FID of the target model is $393.94$, which means that the utility of the target model suffers from a server performance drop.
We leave developing more usable techniques to train a diffusion model with DP-SGD as future work.
Table~\ref{tab:loss_ddpm_dp} in Appendix summarizes the quantitative results of both attacks.

\section{Related work}
\label{sec:Related work}
\noindent
\textbf{Diffusion Models.}
Diffusion Models have attracted increasing attention in the past years.
Sohl-Dickstein et al.~\cite{sohl2015deep} first introduce nonequilibrium thermodynamics to build generative models.
The key idea is to slowly add noise into data in the forward process and learn to generate data from noise through a reverse process.  
Ho et al.~\cite{ho2020denoising} further propose to use parameterization techniques in diffusion models, which enable diffusion models to generate high-quality images. 
Song et al.~\cite{song2019generative} present to train a generative model by estimating gradients of data distribution, i.e. score.
Furthermore, Song et al.~\cite{song2020score} propose a unified framework to describe these diffusion models through the lens of stochastic differential equations.
Beyond image synthesis, diffusion models are also applied to various domains, such as image restoration~\cite{saharia2022palette,wang2022zero}, and text-to-image translation~\cite{ramesh2022hierarchical,nichol2021glide}, even audio and video synthesis~\cite{ho2022video,kong2021diffwave}.  
However, in this work, we study diffusion models from the perspective of privacy.

\smallskip\noindent
\textbf{Membership Inference Attacks.}
There are extensive works on membership inference~(MI) attacks on classification models~\cite{salemml,carlini2021membership,shokri2017membership,ye2021enhanced,liu2022membership}.
Various attack methods under different threat models are proposed, such as using fewer shadow models~\cite{salemml}, using loss values~\cite{carlini2021membership,shokri2017membership,ye2021enhanced,liu2022membership} and using labels of victim models~\cite{choquette2021label,li2021membership}.

In addition to classification models, there are several MI attacks on generative models~\cite{hayes2019logan,hilprecht2019monte,chen2019gan,wu2022membership}.
Hayes et al.~\cite{hayes2019logan} leverage the discriminator of a GAN to mount attacks. 
Chen et al.~\cite{chen2019gan} perform MI attacks by finding a reconstructed sample on the generator of a GAN. 
Nevertheless, all attacks are more specific to GANs and heavily rely on the unique characteristics of GANs, such as discriminators or generators.
They cannot be extended to diffusion models, because diffusion models have different training and sampling mechanisms.
Therefore, our work on membership inference of diffusion models aims to fill this gap.

Another recent work proposed by Somepalli et al.~\cite{somepalli2022diffusion} investigates data replication in diffusion models. 
However, their work is different from our work.
Data replication assumes that adversaries can have the whole training set. Given a generated sample from the diffusion model, they search the training set based on similarity metrics.
If the similarity value is higher than a threshold, it is considered a replication for this generated sample.
In contrast, our work does not assume that adversaries obtain the training set.  
Our work aims to infer whether a training sample is used to train the model, given a diffusion model.

\section{Conclusion}
\label{sec:Conclusion}
In this paper, we have presented the first study about membership inference of diffusion models.
We have developed two types of attack methods: loss-based attack and likelihood-based attack.
We have evaluated our methods on four state-of-the-art diffusion models and two privacy-related datasets~(human faces and medical images).

Our evaluations have demonstrated that diffusion models are vulnerable to membership inference attacks.
To be more specific, our loss-based attack shows that when utilizing loss values from diffusion steps where low levels of noise are added, training samples can be inferred with high true positive rates at low false positive rates, such as $100\%$ TPR  at $0.01\%$ FPR.
Our likelihood-based attack again illustrates that adversaries can achieve the same perfect performance.
Although membership inference becomes more challenging with the increase in the number of training samples, attack performance in the worst-case, i.e. TPR at low FPR, is still significantly higher than random guesses.
Our experimental results on classic privacy protection mechanisms, i.e. diffusion models trained with DP-SGD, further show that DP-SGD alleviates our attacks at the expense of severe model utility. 

Designing an effective differential privacy strategy to produce high-quality images for diffusion models is still a promising and challenging direction.
We will take this as one of our future works.
In addition, it is an interesting direction to study MI attacks of diffusion models in stricter scenarios, such as only obtaining synthetic data.

\vspace{5pt}
\noindent{\bf Acknowledgements:} 
This research was funded in whole by the Luxembourg National Research Fund (FNR), grant reference 13550291.

\bibliographystyle{splncs04}
\bibliography{egbib}

\appendix
\section{Appendix}
In this section, we show additional results and introduce each result in its caption.

\begin{figure*}[!h]
	\centering
	\subfigure[DDPM]{
		\includegraphics[width=0.20\columnwidth]{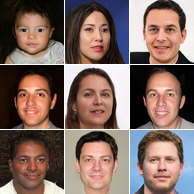}
		\label{fig:generated_ddpm_ffhq1k}
	}
	\subfigure[SMLD]{
		\includegraphics[width=0.20\columnwidth]{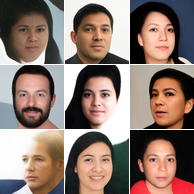}
		\label{fig:generated_smld_ffhq1k}
	}
    \subfigure[VPSDE]{
		\includegraphics[width=0.20\columnwidth]{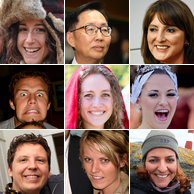}
		\label{fig:generated_vp_cont_ffhq1k}
	}
	\subfigure[VESDE]{
		\includegraphics[width=0.20\columnwidth]{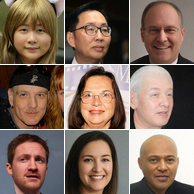}
		\label{fig:generated_ve_cont_ffhq1k}
	}
	
	\caption{Generated images from different target models trained on FFHQ. It is corresponding to Section~\ref{ssec:perf_target_models}.} 
\label{fig:target_gen_imgs}
\end{figure*}

\begin{figure}[!h]
	\centering
			\includegraphics[width=0.2\columnwidth]{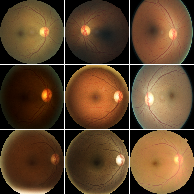}
	\caption{Generated images from the target model SMLD trained on DRD. It is corresponding to Section~\ref{ssec:perf_drd}.}

	\label{fig:generated_smld_drd1k}
\end{figure}

\begin{table}
	\centering
	\caption{Quantitative results of our attacks on SMLD trained on DRD. It is corresponding to Section~\ref{ssec:perf_drd}.}
	\label{tab:loss_ddpm_drd1k_diffdataset}
	\begin{tabular}{cc|rrrrr} 
		\toprule
		Attack           & T   & TPR@     & TPR@     & TPR@     & TPR@      & Accuracy  \\
		&     & 10\%FPR  & 1\%FPR   & 0.1\%FPR & 0.01\%FPR &           \\
		\hline
		Loss-based       & 0   & 7.50\%   & 1.10\%   & 0.00\%   & 0.00\%    & 50.25\%   \\
		& 200 & 11.20\%  & 0.70\%   & 0.10\%   & 0.00\%    & 52.25\%   \\
		& 700 & 80.60\%  & 50.50\%  & 33.34\%  & 18.80\%   & 85.45\%   \\
		& 800 & 93.30\%  & 72.20\%  & 60.00\%  & 40.10\%   & 92.25\%   \\
		& 900 & 79.80\%  & 42.40\%  & 17.70\%  & 12.30\%   & 86.35\%   \\
		& 999 & 43.60\%  & 9.70\%   & 2.00\%   & 0.10\%    & 70.95\%   \\
		\hline\hline
		Likelihood-based & -   & 100.00\% & 100.00\% & 100.00\% & 99.90\%   & 99.95\%   \\
		\bottomrule
	\end{tabular}
\end{table}

\begin{table}[!h]
	\centering
	\caption{Quantitative results of our attacks on DDPM trained with DP-SGD. It is corresponding to Section~\ref{sec:Defenses}.}
	\label{tab:loss_ddpm_dp}
	\renewcommand{\arraystretch}{1.0}
	\scalebox{0.9}{	
	\begin{tabular}{cc|rrrrr} 
		\toprule
		Attacks          & T   & TPR@    & TPR@   & TPR@     & TPR@      & Accuracy  \\
		&     & 10\%FPR & 1\%FPR & 0.1\%FPR & 0.01\%FPR &           \\
		\hline
		Loss-based       & 0   & 8.80\%  & 1.40\% & 0.00\%   & 0.00\%    & 52.25\%   \\
		& 200 & 8.60\%  & 1.40\% & 0.40\%   & 0.30\%    & 53.20\%   \\
		& 500 & 10.70\% & 1.60\% & 0.90\%   & 0.90\%    & 51.85\%   \\
		& 600 & 13.00\% & 2.30\% & 1.00\%   & 1.00\%    & 51.85\%   \\
		& 800 & 11.60\% & 2.10\% & 0.30\%   & 0.30\%    & 51.75\%   \\
		& 999 & 10.40\% & 0.60\% & 0.00\%   & 0.00\%    & 53.90\%   \\
		\hline\hline
		Likelihood-based & -   & 8.40\%  & 1.10\% & 0.20\%   & 0.10\%    & 51.75\%   \\
		\bottomrule
	\end{tabular}
}
\end{table}

\begin{figure*}[!h]
	\centering
	
	\subfigure[FFHQ-1k.]{
		\includegraphics[width=0.30\columnwidth]{fig/loss_ddpm_ffhq1k_TPR.pdf}
		\label{fig:loss_ddpm_ffhq1k_TPR_}
	}	
	\subfigure[FFHQ-10k.]{
		\includegraphics[width=0.30\columnwidth]{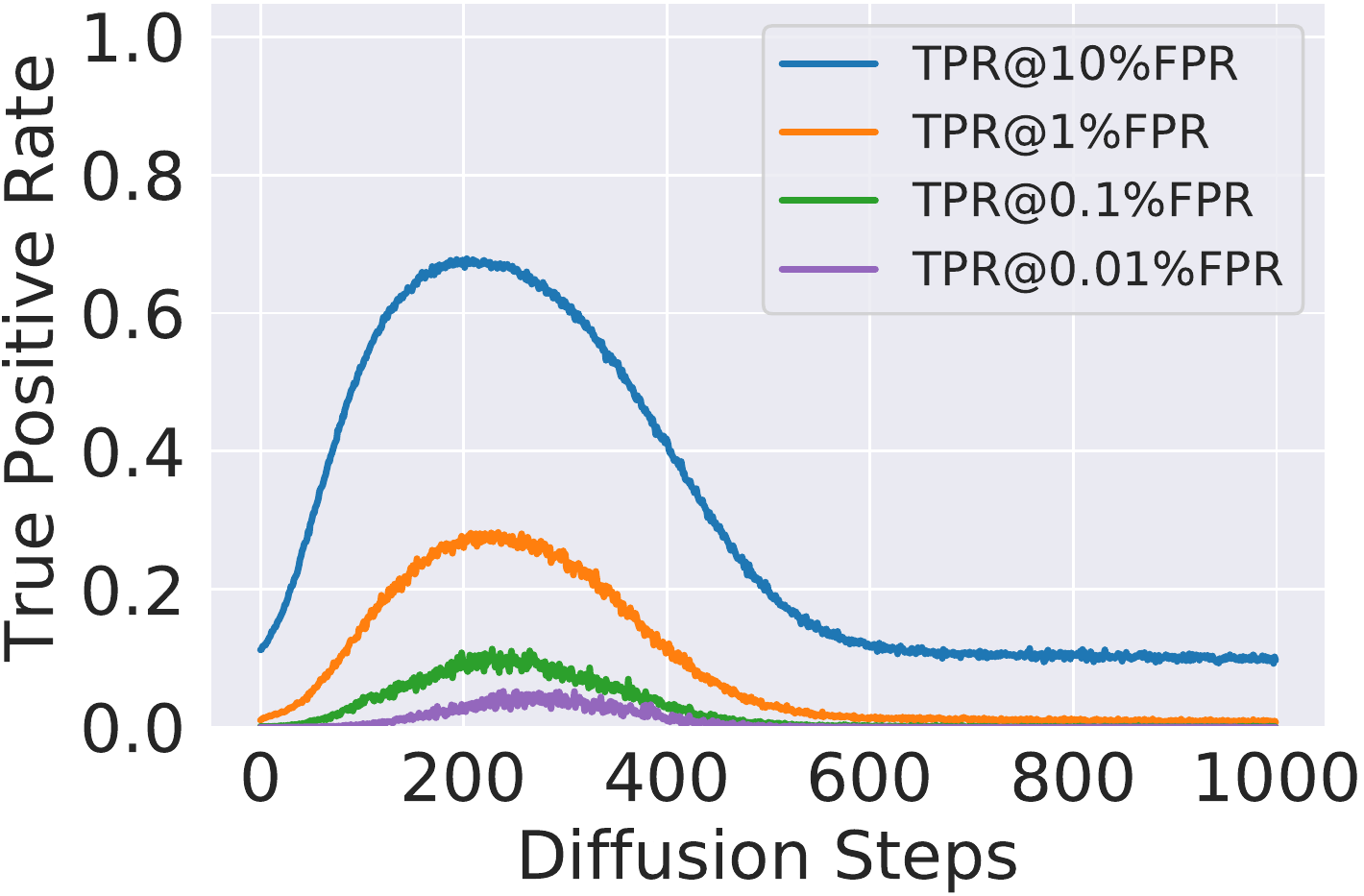}
		\label{fig:loss_ddpm_ffhq10k_TPR}
	}
	\subfigure[FFHQ-30k.]{
		\includegraphics[width=0.30\columnwidth]{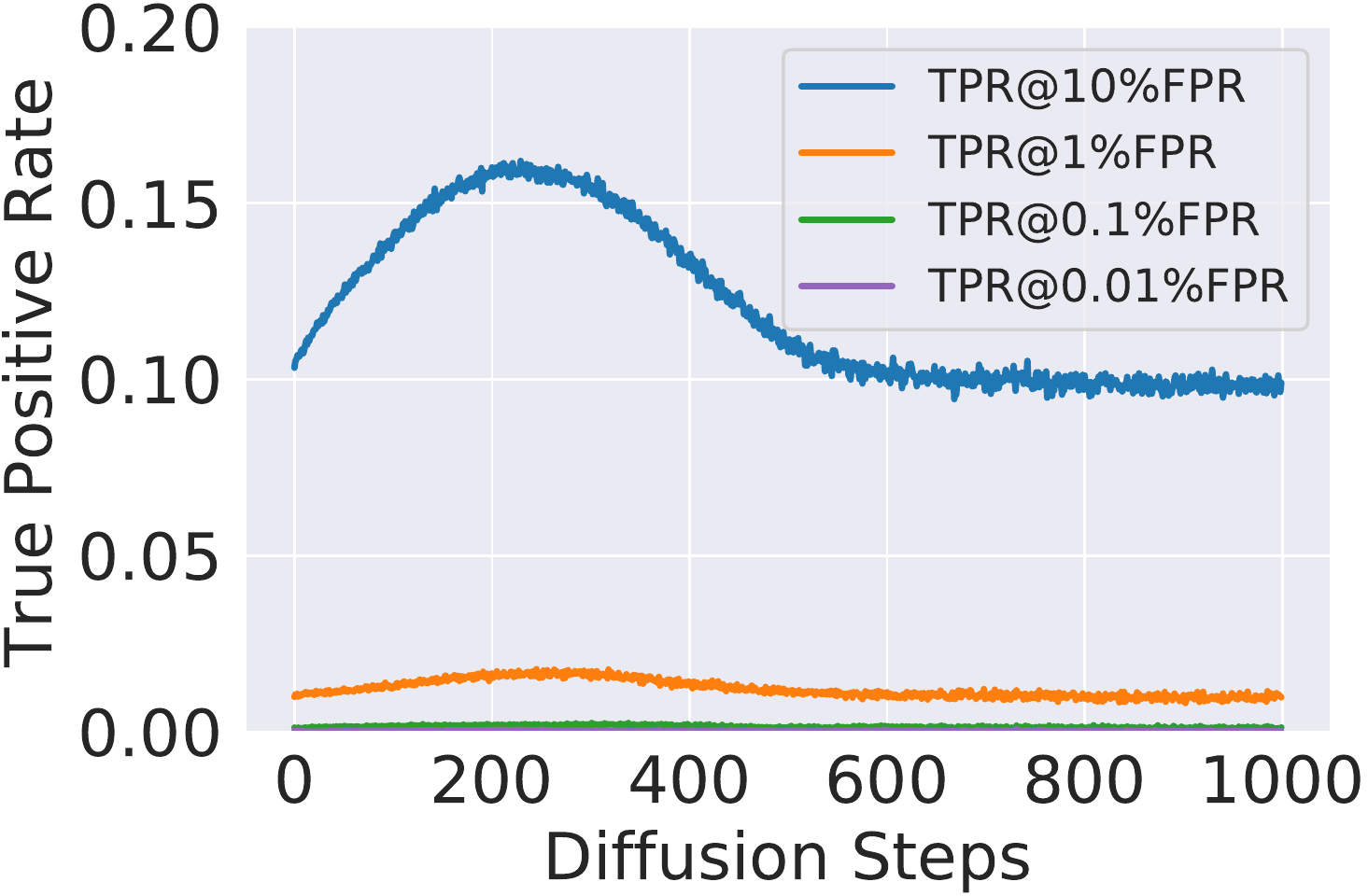}
		\label{fig:loss_ddpm_ffhq30k_TPR}
	}
		
	\caption{Performance of loss-based attacks with different sizes of a training set. The target model is DDPM trained on FFHQ. It is corresponding to Section~\ref{ssec:effects_size_data}.}
	\label{fig:loss_ddpm_ffhq_diffsize}
\end{figure*}

\end{document}